\documentclass{emulateapj}

\usepackage{amsmath,amssymb,bm,url}
\usepackage{graphics,graphicx}

\newcommand{\beq}{\begin{equation}}
\newcommand{\eeq}{\end{equation}}

\newcommand{\bftheta}{\bm \theta}
\newcommand{\bfalpha}{\bm \alpha}

\shorttitle{Probing the solar interior through GW lensing}
\shortauthors{Takahashi, Morisaki \& Suyama}

\begin{document}

\title{Probing the solar interior with lensed gravitational waves from known pulsars}


\author{Ryuichi Takahashi$^1$, Soichiro Morisaki$^2$ and Teruaki Suyama$^3$}
\affil{$^1$ Faculty of Science and Technology, Hirosaki University, 3 Bunkyo-cho, Hirosaki, Aomori 036-8561, Japan \\
$^2$ Institute for Cosmic Ray Research (ICRR), KAGRA Observatory, The University of Tokyo, Kashiwa City, Chiba 277-8582, Japan \\
$^3$ Department of Physics, Tokyo Institute of Technology, 2-12-1 Ookayama, Meguro-ku, Tokyo 152-8551, Japan}


\begin{abstract}

When gravitational waves (GWs) from a spinning neutron star arrive from behind the Sun, they are subjected to gravitational lensing that imprints a frequency-dependent modulation on the waveform.
This modulation traces the projected solar density and gravitational potential along the path as the Sun passes in front of the neutron star.
We calculate how accurately the solar density profile can be extracted from the lensed GWs using a Fisher analysis.
For this purpose, we selected three promising candidates (the highly spinning pulsars J1022+1001, J1730-2304, and J1745-23) from the pulsar catalog of the Australia Telescope National Facility. 
The lensing signature can be measured with $3 \sigma$ confidence when the signal-to-noise ratio (SNR) of the GW detection reaches $100 \, (f/300 {\rm Hz})^{-1}$ over a one-year observation period (where $f$ is the GW frequency).
The solar density profile can be plotted as a function of radius when the SNR improves to $\gtrsim 10^4$. 

\end{abstract}


\section{Introduction} \label{sec:intro}

Since the direct detection of a gravitational-wave (GW) signal from a merging black hole binary~\citep[GW150914;][]{Abbott2016}, GW astronomy has attracted increasing interest.
Currently, all GW sources detected by the ground-based detectors of the Laser Interferometer Gravitational-Wave Observatory (LIGO), the Virgo interferometer, and the Kamioka Gravitational Wave Detector (KAGRA) are compact binary coalescences (CBCs) of stellar-mass black holes and neutron stars.
The LIGO-Virgo-KAGRA (LVK) collaboration reported $\sim 90$ candidates of CBC events~\citep{LVK2021}. 
However, the ground-based detectors are also expected to detect spinning neutron stars~\citep[e.g.,][]{Abbott2022}.
A neutron star that is non-axisymmetric around its spin axis, resulting in a so-called mountainous profile, emits continuous GWs.
Non-axisymmetry may be caused by crustal deformation, magnetic fields, and mass accretion from the star's companion (e.g., see recent reviews by \cite{GG2018} and \cite{Riles2022}).
Although the LVK collaboration has been searching for continuous GW signals, no event has yet been reported~\citep{Abbott2021,Abbott2022,Abbott2022b,LVK2022}.

If a GW signal encounters the Sun along its path, gravitational lensing imprints a frequency-dependent modulation on the waveform.
In geometrical optics (i.e., the zero-wavelength limit of GWs), the solar density modulates the amplitude with a magnification effect, while the gravitational potential modulates the phase by imposing a potential (or Shapiro) time delay.
These modulations can be obtained along the transversal path of the Sun moving in front of the source (the duration of this movement is approximately half a day).
Therefore, in principle, one can probe the solar interior using the lensed signal.
GW lensing by the Sun has been studied as a tool for amplifying the strain amplitude of GWs from a distant source and probing the solar structure~\citep[e.g.,][]{CL1974,Sonnabend1979,PN2008,Marchant2020}. 
Before the 1980s, these studies were based on geometrical optics. 
\cite{BH1981} first demonstrated that the diffraction effect caused by the finite wavelength of GWs suppresses magnification near the focal point at lower frequencies ($f \lesssim 10^4 \, {\rm Hz}$).
Recently, \cite{Marchant2020} proposed that the solar structure can be probed through the lensed GW signal from a pulsar behind the Sun.
However, they only roughly estimated the detectability of the lensing signature, without calculating the measurement accuracy of the density profile.
Very recently, \cite{Jung2022} reported that the Fresnel scale is comparable to the solar radius within the frequency band of the ground-based detectors; therefore, the density profile can (in principle) be probed with the lensing modulation on a chirp signal from a CBC. 
 
In this study, we investigate the accuracy of measuring the solar density profile using the lensed signals of known pulsars.
Because ground-based detectors detect GW wavelengths longer than (or comparable to) the solar Schwarzschild radius, gravitational lensing should employ wave optics~\citep[e.g.,][]{Ohanian1974,BM1975,sef92,ND1999,TN2003,Dai2018,Oguri2019,Suvorov2022,Liao2022}.
We first calculate the lensed waveform based on wave optics. 
We discuss the effects of frequency and impact parameter on the waveform (Section \ref{sec:amplf}).
Then we extract the known pulsars crossing behind the Sun 
from the Australia Telescope National Facility (ATNF) pulsar catalog~\citep{ATNF_pulsar_catalog}.
From the extracted list, we select suitable candidates by calculating the lensing modulations of these pulsars (Section \ref{sec:pulsars}).
Using these candidates, we calculate the detectability of the lensing signature and the accuracy of measuring the solar density profile through a Fisher analysis (Section \ref{sec:fisher}). 
We roughly estimate the number of Galactic millisecond pulsars (MSPs) behind the Sun, which are potentially detectable by near-future radio surveys (Section \ref{sec:MSPs}).  
We discuss the lensing by a Galactic star, which may be confused with the lensing by the Sun (Section \ref{sec:discussion}).
The study findings are summarized in Section \ref{sec:conclusion}.

\section{Amplitude and phase modulations of solar lensing} \label{sec:amplf}

This section introduces the amplitude and phase modulations of a lensed waveform, based on wave optics.

\subsection{Lensed waveform}

\begin{figure}
\epsscale{1.1}
\plotone{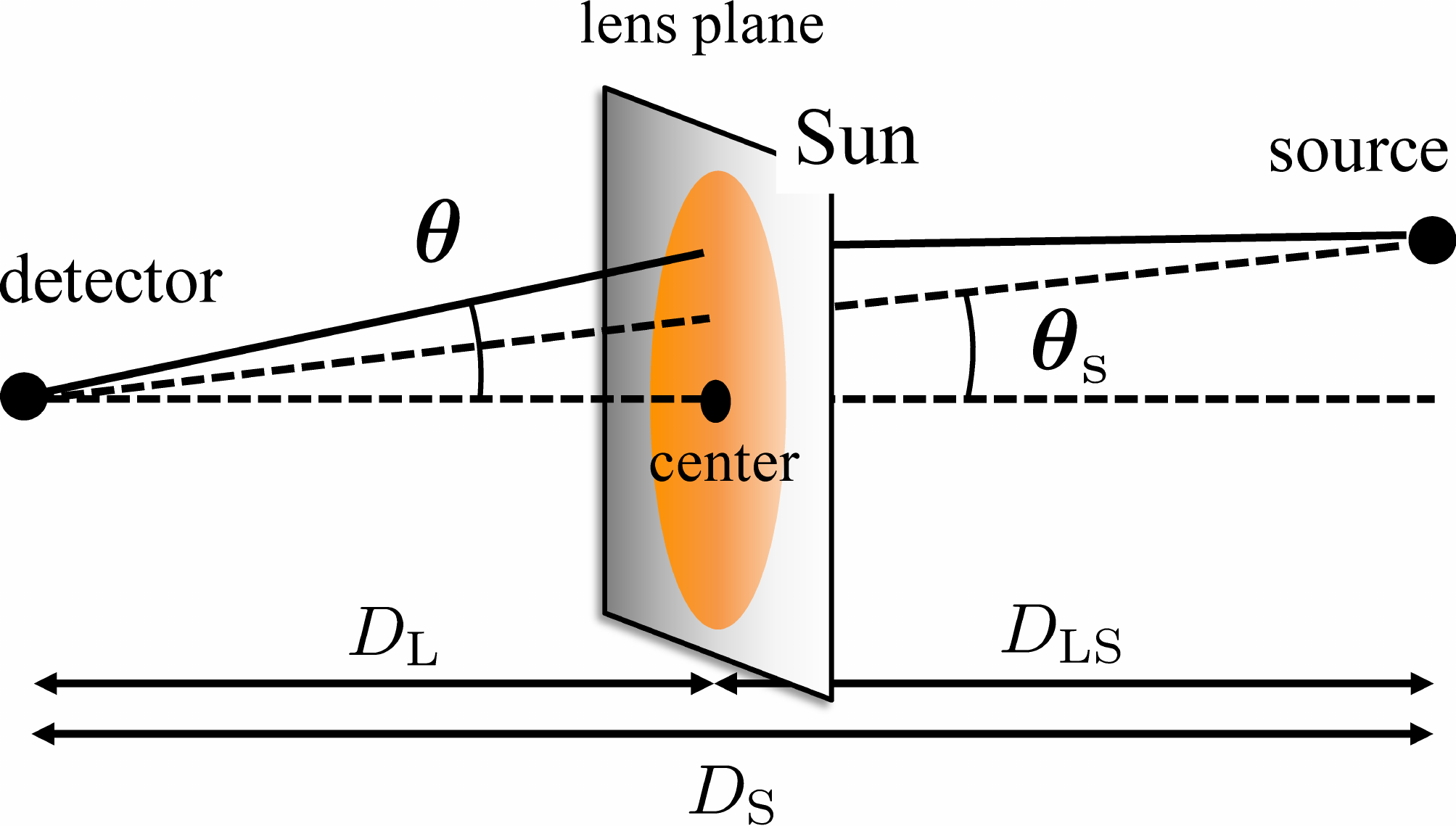}  
\caption{Configuration of the detector, Sun, and source.
$D_{\rm L}$,  $D_{\rm LS}$, and $D_{\rm S}$ denote the distances between them. Solid line is the GW path. Angular coordinates $\bftheta$ and $\bftheta_{\rm s}$ describe the incoming direction of the GWs and source position, respectively, with respect to the solar center.} 
\label{fig_lens_config}
\vspace*{0.3cm}
\end{figure}

Figure \ref{fig_lens_config} shows the configuration of the detector, Sun, and source.
The solid line describes a single GW path\footnote{ 
In wave optics, the lensed waveform is the superposition of many GW paths from the source to the detector (Eq.~(\ref{ampf})). The solid line represents one of them.}.
We assume the thin-lens approximation, in which the solar density and its gravitational potential are projected onto a plane perpendicular to the line of sight.
The thin-lens approximation is valid within an error less than $0.5 \, \%$ ($\simeq R_\odot/D_{\rm L}$) on the lensed waveform~\cite[Appendix \ref{sec:systematics};][]{Suyama2005}.
Furthermore, we assume the flat-sky approximation, in which $|\bftheta|,|\bftheta_{\rm s}| \ll 1$.
As the coordinate system is fixed at the solar position, the background source (pulsar) moves along the ecliptic longitude, following the annual solar motion.
The distance to the Sun is $D_{\rm L}=1 \, {\rm au}$ and the solar angular radius is $\theta_\odot = \arctan[ R_\odot/(1 \, {\rm au}) ] \simeq 16$ arcmin. 
The annual modulation of $D_{\rm L}$ is ignored because it changes by $< 2  \, \%$ as the Earth moves elliptically around the Sun.

The lensed waveform $\tilde{h}^{\rm L}(f)$ in the frequency domain is obtained by multiplying the unlensed waveform $\tilde{h}(f)$ by a function $F(f;\bftheta_{\rm s})$~\cite[e.g.,][]{Nakamura1998,TN2003}:
\begin{align}
 \tilde{h}^{\rm L}(f;\bftheta_{\rm s}) &= F(f;\bftheta_{\rm s}) \tilde{h}(f),  \nonumber \\ 
  &= \left[ 1+A_F(f;\bftheta_{\rm s}) \right] {\rm e}^{{\rm i} \varPhi_F(f;\bftheta_{\rm s})} \tilde{h}(f). 
  \label{AF_PhiF}
\end{align}
In the second line, $A_F \, (\equiv |F|-1)$ and $\varPhi_F \, (\equiv \ln (F/|F|))$ represent the amplitude and phase modulations, respectively. 
The GW polarization is ignored because the polarization rotation due to lensing is negligibly small~\citep[e.g.,][]{Hou2019,Ezquiaga2021,Dalang2022}. 
The function $F$, called the amplification factor or transmission factor, is given by the following diffraction integral \citep[e.g.,][]{sef92}:
\beq
F(f;\bftheta_{\rm s}) = \frac{D_{\rm L} D_{\rm S}}{c D_{\rm LS}} \frac{f}{\rm i}
\int \! {\rm d}^2 \theta \exp \left[ 2 \pi {\rm i} f t_{\rm d}(\bftheta,\bftheta_{\rm s}) \right],
\label{ampf}
\eeq
where the time delay is given by
\beq
 t_{\rm d}(\bftheta,\bftheta_{\rm s})= \frac{1}{c}
 \left[ \frac{D_{\rm L} D_{\rm S}}{2 D_{\rm LS}} \left| \bftheta-\bftheta_{\rm s} \right|^2
   - \frac{\hat{\psi}(\bftheta)}{c^2} \right].
\label{time_delay}
\eeq
The first and second terms are the geometrical and potential (or Shapiro) time delays, respectively.
Because the distance to the Sun is much shorter than the distance to the source, hereafter we assume $D_{\rm S}/D_{\rm LS} \simeq 1$. 
The two-dimensional lens potential $\hat{\psi}(\bftheta)$ is determined from the solar-projected density profile $\Sigma(\bftheta)$ using the Poisson equation:
\begin{equation}
 \nabla^2_\theta \hat{\psi}(\bftheta) = 8 \pi G D^2_{\rm L} \Sigma(\bftheta).
 \label{Poisson}
\end{equation}
Outside the Sun, the potential is identical to that of a point mass:
\beq
  \hat{\psi}(\theta) = 4 G M_\odot \ln \left( \frac{\theta}{\theta_{\rm E}} \right) ~~{\rm for}~ \theta>\theta_\odot,
\label{psi_pointmass}
\eeq
where $\theta_{\rm E}$ is an arbitrary constant, here set to the angular Einstein radius; i.e., $\theta_{\rm E}=[4 G M_\odot/(c^2 D_{\rm L})]^{1/2} \simeq 0.043 \, \theta_\odot$.
The Sun is modeled using the BS05(OP) spherical density model\footnote{The numerical table can be downloaded from \url{http://www.sns.ias.edu/~jnb/SNdata/sndata.html\#bs2005}.} in \cite{Bahcall:2005}.
The model density agrees within $2 \, \%$ of the helioseismological results across the whole radius (their Fig.~1).
It also agrees within $3 \, \%$ and $4 \, \%$ of the recent solar density models of B16-GS98 and B16-AGSS09met, respectively, in \cite{Vinyoles2017}.
As the density profile is spherical, the integration (\ref{ampf}) can be performed over the azimuth and 
$F$ reduces to
\begin{align}
 F(f;\theta_{\rm s}) &= \frac{2 \pi f D_{\rm L}}{{\rm i} c} \int_0^\infty \!\! {\rm d} \theta \, \theta J_0 \left( \frac{2 \pi f D_{\rm L}}{c} \theta \theta_{\rm s} \right) \nonumber \\
 & \times \exp \left[ 2 \pi {\rm i} f \left\{ \frac{1}{2 c} D_{\rm L} \left( \theta^2 + \theta_{\rm s}^2 \right) - \frac{\hat{\psi}(\theta)}{c^3} \right\} \right],
\label{ampf2}
\end{align}
where $J_0$ is the zero-order Bessel function.  
Equation (\ref{ampf2}) is numerically integrated using integration by parts (Appendix A of \cite{RT2004}; \cite{GL2020}).

Let us comment on the validity of the spherical density model of the Sun. 
The spin rotation makes the equatorial radius larger than the polar radius, but the observed relative difference between these surface radii is very small $\approx 10^{-5}$~\cite[Chapter 7 of][]{Stix2004}. 
The quadrupole mass moment of the Sun ($J_2$)\footnote{Modeling the Sun as a spheroidal body about its spin axis ($z$ axis), $J_2$ is simply $J_2=(I_{zz}-I_{xx})/(M_\odot R_\odot^2)$ with the moment of inertia $I_{ij}$.  A positive (negative) $J_2$ means an oblate (prolate) shape.} generates the precession of Mercury's orbital perihelion (which is much smaller than the known general-relativistic effect).
The precession was precisely measured by the MESSENGER (MErcury Surface, Space ENviroment, GEochemistry, and Ranging) spacecraft\footnote{\url{https://solarsystem.nasa.gov/missions/messenger/in-depth/}}, which gave a constraint on $J_2 \simeq 2 \times 10^{-7}$~\citep{Park2017}. These observations suggest that $F$ for the spherical model would be valid with an error of $\lesssim 10^{-5}$.

\subsection{High- and low-frequency limits}

\begin{figure*}
\epsscale{1.1}
\vspace*{-4cm}
\plotone{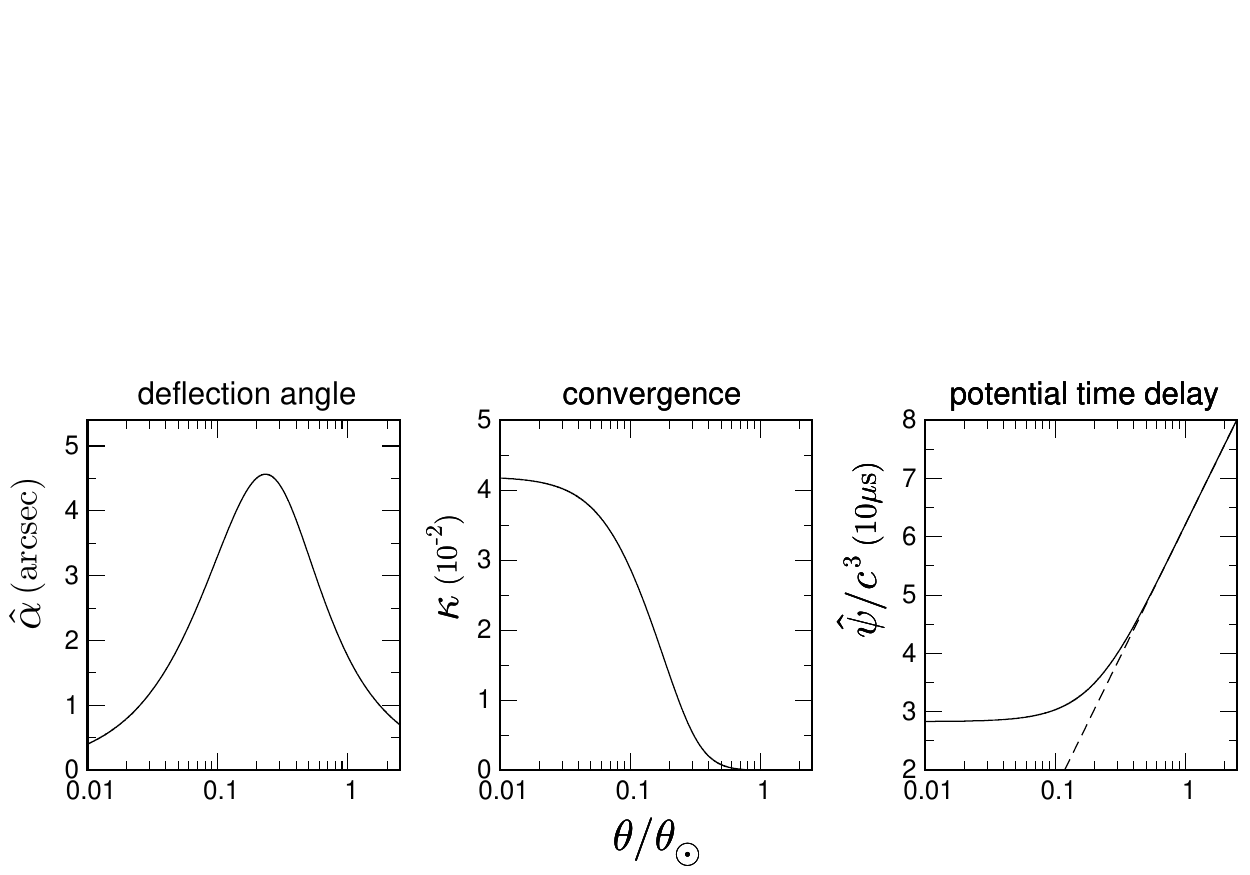}     
\caption{Solar deflection angle, convergence, and potential (or Shapiro) time delay as functions of angular radius $\theta$ (normalized by the solar radius $\theta_\odot$). The convergence $\kappa$ in the middle panel is the dimensionless projected density profile. The potential time delay in the right panel is the projected gravitational potential divided by $c^3$. Near and outside the surface ($\theta/\theta_\odot \gtrsim 1$), the result approaches that of Eq.~(\ref{psi_pointmass}) for a point mass  
(the dashed line).} 
\label{fig_alpha-conv-psis}
\vspace*{0.8cm}
\end{figure*}

This subsection presents the high- and low-frequency limits of the amplification factor.
In the high-frequency limit (i.e., the geometrical-optics limit), because the exponential term of $F$ in Eq.~(\ref{ampf}) oscillates violently, a stationary point of $t_{\rm d}(\bftheta,\bftheta_{\rm s})$ contributes to the integral~\citep[e.g.,][]{sef92}.
This point is a solution of $\nabla_\theta t_{\rm d}(\bftheta,\bftheta_{\rm s})=0$ :
\beq
  \bftheta_{\rm s} = \bftheta - \hat{\bfalpha}(\bftheta),
\label{lens_eq}
\eeq
from Eq.~(\ref{time_delay}), where $\hat{\bfalpha}=\nabla_\theta \hat{\psi} / (c^2 D_{\rm L})$ is the deflection angle.
Equation (\ref{lens_eq}) is the lens equation.
The image position $\bftheta_{\rm i}$ is obtained by solving Eq.~(\ref{lens_eq}).
Note that only a single image is formed irrespective of $\bftheta_{\rm s}$.
In the high-frequency limit, $F$ reduces to the following simple form:
\beq
  \lim_{f \rightarrow \infty} F(f;\bftheta_{\rm s}) = \left| \mu(\bftheta_{\rm i},\bftheta_{\rm s}) \right|^{1/2} \exp \left[ 2 \pi i f t_{\rm d}(\bftheta_{\rm i},\bftheta_{\rm s}) \right],
\eeq
where $\mu =[\,\det (\partial \bftheta_{\rm s} / \partial \bftheta_{\rm i})\,]^{-1}$ is the magnification of the image.

In the weak-gravitational-field limit, 
the magnification can be approximated as $\mu \simeq 1+2 \kappa$, where $\kappa = \Sigma/\Sigma_{\rm cr}$ is the convergence and $\Sigma_{\rm cr}=c^2/(4 \pi G D_{\rm L})$ is the critical density~\citep[e.g.,][]{BS2001}.
Accordingly, $A_F$ and $\varPhi_F$ reduce to
\begin{align}
  \lim_{f \rightarrow \infty} \! A_F(f;\bftheta_{\rm s}) &\simeq \kappa(\bftheta_{\rm s}) = \frac{4 \pi G D_{\rm L}}{c^2} \Sigma(\bftheta_{\rm s}), \nonumber \\
  \lim_{f \rightarrow \infty} \! \varPhi_F(f;\bftheta_{\rm s}) &= 2 \pi f t_{\rm d}(\bftheta_{\rm i},\bftheta_{\rm s}) \simeq - 2 \pi f \frac{\hat{\psi}(\bftheta_{\rm s})}{c^3}.
\label{AP_geo}  
\end{align}
Therefore, $A_F$ and $\varPhi_F$ trace the projected density profile and gravitational potential, respectively.   

Figure \ref{fig_alpha-conv-psis} plots the deflection angle, convergence, and gravitational potential as functions of $\theta$ (as these variables are functions of $\theta$, we plot $\theta$ rather than $\theta_{\rm s}$ along the x-axis). 
The deflection angle is consistent with the known result $\hat{\alpha}=4GM_\odot/(c^2 R_\odot) \simeq 1.75\,{\rm arcsec}$ at the surface. 
The maximum deflection is $\hat{\alpha} \simeq 4.57\,{\rm arcsec}$ at $\theta \simeq 0.23 \, \theta_\odot$.
Because $\hat{\alpha}$ is much smaller than the solar radius ($\theta_\odot \simeq 16\,{\rm arcmin}$), we can safely set $\bftheta_{\rm i} \simeq \bftheta_{\rm s}$ from Eq.~(\ref{lens_eq}). 
Therefore, the second term of $t_{\rm d}$ in Eq.~(\ref{time_delay}) exceeds the first term in the geometrical-optics limit.
Meanwhile, the convergence $\kappa$ is maximized at $\sim 0.04$ near the center.
The profile is flat at the core ($\theta \lesssim 0.03 \, \theta_\odot$), but it drops steeply at $\theta \gtrsim 0.1 \, \theta_\odot$.
The typical gravitational time delay is $\approx 10^{-5}$ s. 
Near the center, the potential in Eq.~(\ref{Poisson}) can be approximated as $\hat{\psi}(\theta)$$={\rm const.}$$+ (c^2 D_{\rm L}/2) \, \kappa(\theta\!=\!0) \, \theta^2$$+ \mathcal{O}(\theta^3)$. 
The results of Fig.~\ref{fig_alpha-conv-psis} are consistent with previous works on lensing by the transparent Sun~\citep{BH1981,PN2008,Marchant2020,Jung2022}.

In the large-angle limit $\theta_{\rm s} \gg \theta_\odot$, the geometrical-optics result (\ref{AP_geo}) is recovered because $2  \pi f t_{\rm d} \gg 1$.  
As the gravitational potential in Eq.~(\ref{psi_pointmass}) is valid only for $\theta_\odot < \theta \ll 1$, it must be replaced with~\citep[e.g.,][]{BH1986}
\beq
 \hat{\psi}(\theta) = 2 GM_\odot \ln \left\{ \frac{2 (1-\cos \theta )}{\theta_{\rm E}^2} \right\},
\label{psi_largetheta}
\eeq
when $\theta \gtrsim 1$.

We now investigate the low-frequency limit.
Changing the variables to $\bftheta^\prime=\sqrt{f} \bftheta$ and $\bftheta_{\rm s}^\prime=\sqrt{f} \bftheta_{\rm s}$ in Eq.~(\ref{ampf}), $F$ is rewritten as
\begin{align}
   F(f;\bftheta_{\rm s}) &= \frac{D_{\rm L}}{{\rm i} c} \int \! {\rm d}^2 \bftheta^\prime \exp \left[ \frac{2 \pi {\rm i}}{c} \left\{ \frac{D_{\rm L}}{2} \left| \bftheta^\prime-\bftheta^\prime_{\rm s} \right|^2
   \right. \right.
  \nonumber \\
   & ~~~~\left. \left. - \frac{f}{c^2} \hat{\psi} \left( \frac{\bftheta^\prime}{\sqrt{f}} \right) \right\} \right].
   \label{ampf3}
\end{align}
In the low-frequency limit, $\hat{\psi}$ can be replaced with its value at infinity:
\beq
\lim_{f \rightarrow 0} \hat{\psi} \left( \frac{\theta^\prime}{\sqrt{f}} \right) = 4 G M_\odot \ln \left( \frac{\theta^\prime}{\sqrt{f} \theta_{\rm E}} \right),
\notag
\eeq
from Eq.~(\ref{psi_pointmass}). 
Inserting this $\hat{\psi}$ into Eq.~(\ref{ampf3}) and expanding $f$ as a Taylor series, we have
\begin{align}
  \lim_{f \rightarrow 0} A_F(f;\bftheta_{\rm s}) &= \frac{2 \pi^2 G M_\odot}{c^3} f, \nonumber \\
  \lim_{f \rightarrow 0} \varPhi_F(f;\bftheta_{\rm s}) &= \frac{4 \pi G M_\odot}{c^3} f \left[ \gamma + \ln \left( \frac{4 \pi G M_\odot}{c^3} f \right) \right],
\label{AP_lowf}  
\end{align}
where $\gamma=0.5772\cdots$ is Euler's constant.

\subsection{Amplitude and phase modulations} 
\label{sec:ap_modulation}

\begin{figure*}
\vspace*{0cm}
\begin{minipage}{\columnwidth}
\includegraphics[width=1.1\columnwidth]{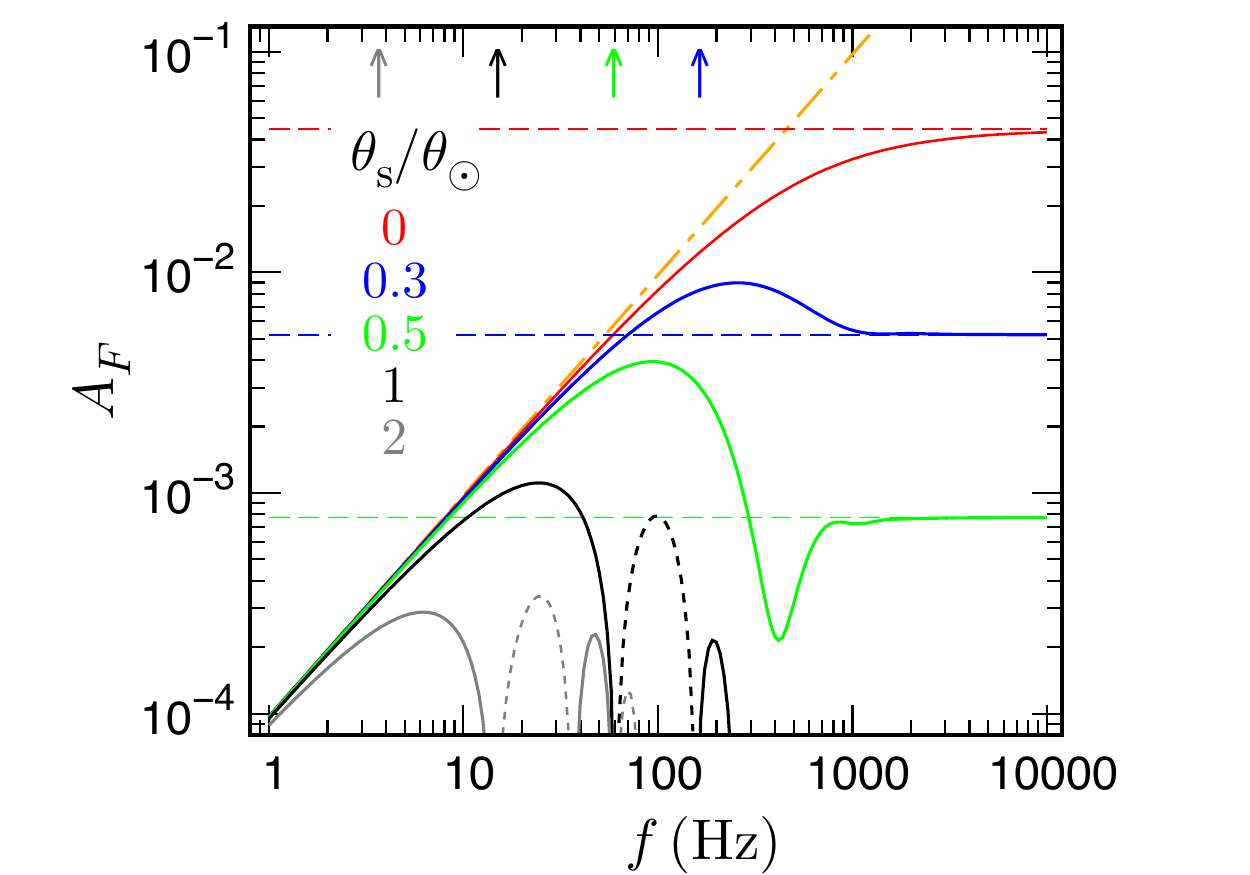}  
\end{minipage}
\begin{minipage}{\columnwidth}
\includegraphics[width=1.1\columnwidth]{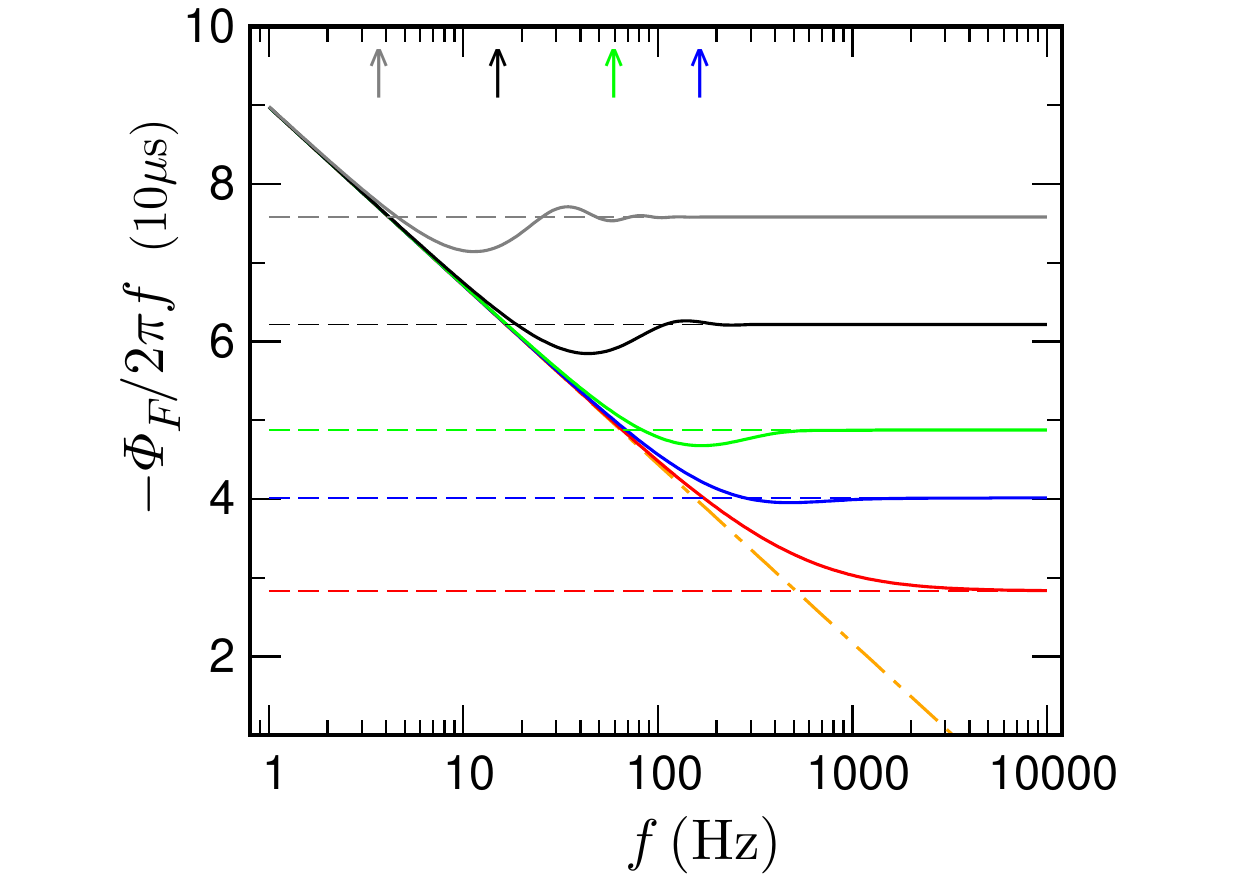}  
\end{minipage}
\caption{Amplitude and phase modulations as functions of frequency at various source positions $\theta_{\rm s}/\theta_\odot=0$--$2$.
Horizontal dashed lines and dot-dashed orange line represent the analytical results in the high- and low-frequency limits, respectively, given by Eqs.~(\ref{AP_geo}) and (\ref{AP_lowf}).
Up arrows indicate the rough boundary between the wave- and geometrical-optics regions for each $\theta_{\rm s}$ (see the description after Eq.~(\ref{Fresnel}) in the text). 
In the left panel, dotted curves denote negative values ($A_F<0$).
In the right panel, the phase modulation is divided by $2 \pi f$, which corresponds to an arrival time delay.
}
\vspace*{0.3cm}
\label{fig_ampf}
\end{figure*}

\begin{figure*}
\vspace*{0cm}
\begin{minipage}{\columnwidth}
\includegraphics[width=1.1\columnwidth]{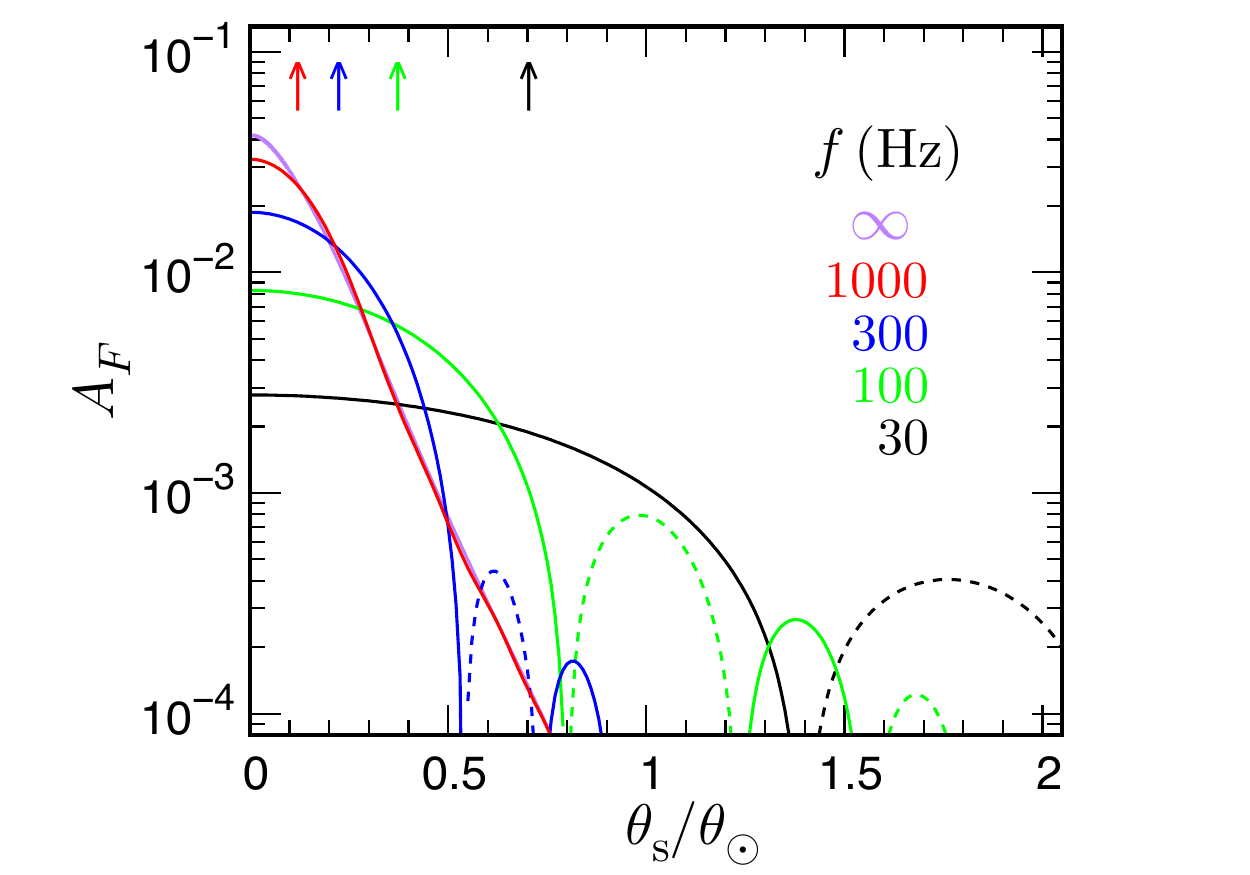}   
\end{minipage}
\begin{minipage}{\columnwidth}
\includegraphics[width=1.1\columnwidth]{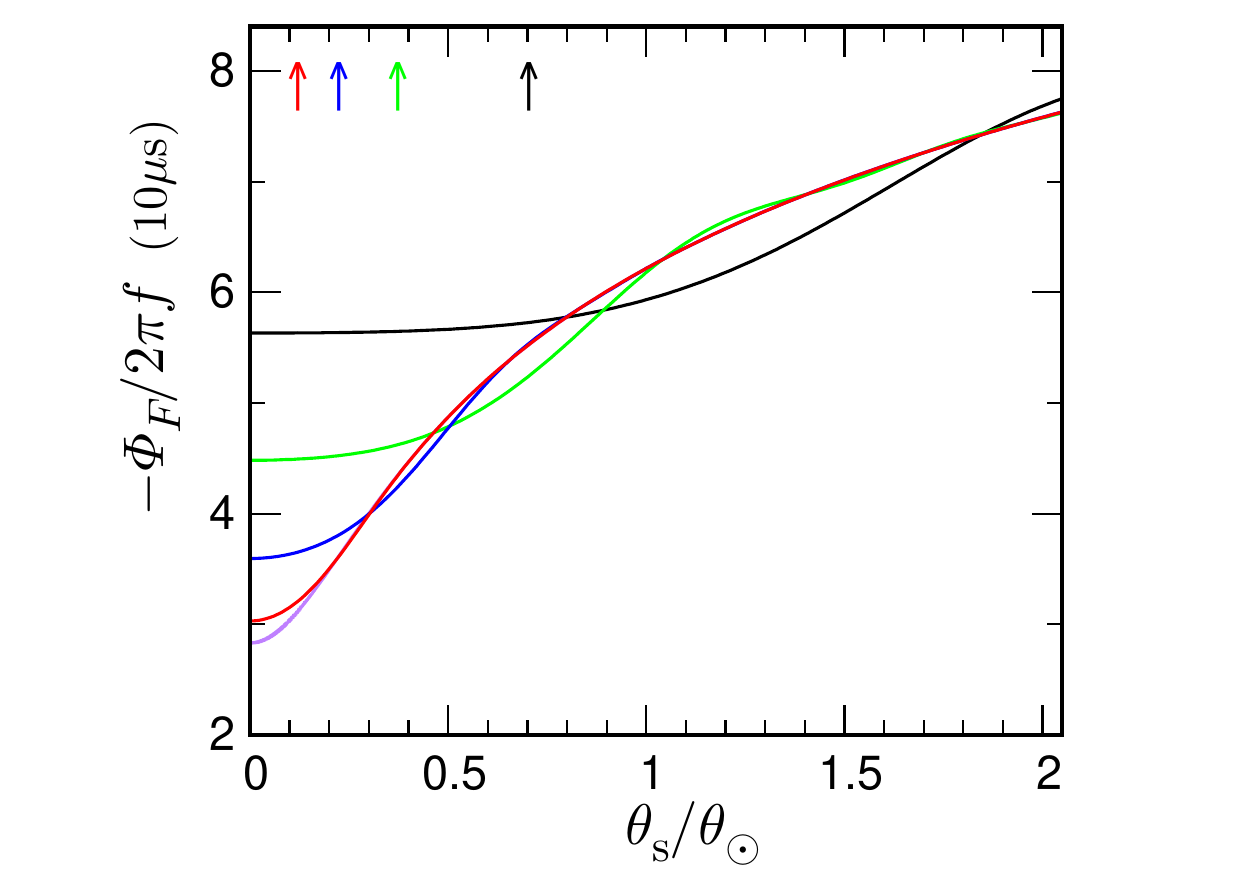}   
\end{minipage}
\caption{Same as Fig.\ref{fig_ampf}, but plotted as functions of $\theta_{\rm s}/\theta_\odot$ for various $f$. Red curve overlaps the purple curve everywhere except near the center ($\theta_{\rm s}/\theta_\odot \lesssim 0.2$).}
\vspace*{0.3cm}
\label{fig_ampf-of-y}
\end{figure*}

Figure \ref{fig_ampf} plots $A_F$ and $\varPhi_F$ as functions of $f$ for various $\theta_{\rm s}$. 
In the high- and low-frequency limits, the plots are consistent with the analytical results of Eqs.~(\ref{AP_geo}) and (\ref{AP_lowf}), respectively.
As the frequency of a chirp signal from an inspiral binary  sweeps from low to high, $A_F$ increases proportionally to $f$ in the wave-optics regime but approaches a constant value of $\kappa$ in the geometrical-optics regime.
Similarly, $-\varPhi_F/(2 \pi f)$ decreases logarithmically, but approaches a constant value of the potential time delay ($\hat{\psi}/c^3$).
These modulations are observed even outside the Sun ($\theta_{\rm s}=2 \theta_\odot$).
Therefore, within the frequency range of ground-based detectors, the solar lensing imprints unique frequency-dependent modulations on both the amplitude and phase.
The solar structure can (in principle) be extracted from the chirp signal~\citep[see the recent study by][]{Jung2022}.
However, in the high-frequency limit, solar structure extraction from the signal (even from a chirp signal) is impossible at a single $\theta_{\rm s}$ because the constant $A_F$ is degenerate with an intrinsic amplitude of GWs. 
Similarly, the constants $\varPhi_F$ and $\varPhi_F/(2 \pi f)$ are degenerate with an intrinsic phase and arrival time, respectively.
Therefore, in geometrical optics, the lensing modulations are indistinguishable from the intrinsic source properties (unless lensed signals with various $\theta_{\rm s}$ from a single source are available).
Figure \ref{fig_ampf} is consistent with Fig.~2 of \cite{Jung2022}.

To discuss the behavior of these modulations in more detail, let us introduce the angular Fresnel scale~\citep{Macquart2004,RT2006}:
\begin{align}
    \theta_{\rm F} &= \left( \frac{c}{2 \pi f} \frac{1}{D_{\rm L}} \right)^{1/2}, \notag \\
    &\simeq 0.38 \, \theta_\odot \left( \frac{f}{100 \, {\rm Hz}} \right)^{-1/2}.
    \label{Fresnel}
\end{align}
In the low-frequency limit, the lensed signal probes a circle of radius $\theta_{\rm F}$ around the source position~\citep[Section 2 and Fig. 1 of][]{Choi2021}.
In other words, the Fresnel scale can be interpreted as the effective source radius~\citep{Oguri2020}. 
The effective source radius shrinks with increasing $f$ in the chirp signal. 
In the high-frequency limit, the lensing probes a small region around the image position.
As the modulations at different $f$ can probe different regions, one can in principle probe the density profile~\citep{Jung2022}.
Here, $\theta_{\rm F}=\theta_{\rm s}$ (unless $\theta_{\rm s}=0$) roughly represents the boundary between the wave- and geometrical-optics regions.
This boundary condition is derived from $2\pi f t_{\rm d}=1$ with neglecting the potential term in Eq.~(\ref{time_delay}) ~(Section II.B of \cite{Choi2021} and Section 3 of \cite{Jung2022})\footnote{This condition can also be derived from the intersection of the low- and high-frequency limits of the phase modulation. Setting $\lim_{f \rightarrow 0} \varPhi_F = \lim_{f \rightarrow \infty} \varPhi_F$ in Eqs.~(\ref{AP_geo}) and (\ref{AP_lowf}) with the point-mass potential (\ref{psi_pointmass}) (which is valid only for $\theta_{\rm s} \gtrsim 0.3 \, \theta_\odot$ in the right panel of Fig.~\ref{fig_alpha-conv-psis}), one obtains $\theta_{\rm F} \simeq 0.94 \, \theta_{\rm s}$.}.
The boundary is plotted as a series of up arrows in Figs.~\ref{fig_ampf} and \ref{fig_ampf-of-y}.

Figure \ref{fig_ampf-of-y} plots $A_F$ and $\varPhi_F$ as functions of $\theta_{\rm s}$ for various fixed $f$.
As the GWs from a pulsar are continuous, one can (in principle) measure $A_F$ and $\Phi_F$ as functions of $\theta_{\rm s}$ for a pulsar moving behind the Sun. 
For smaller (larger) $\theta_{\rm s}$, the exponential term of the amplification factor ($=2 \pi f t_{\rm d}$) is smaller (larger) and the results approach the low-frequency (high-frequency) limit in Eq.~(\ref{AP_lowf}) (Eq.~(\ref{AP_geo})). 
At $f=1000$ Hz, the result almost matches that of geometrical optics, except in the near-center region ($\theta_{\rm s}/\theta_\odot \lesssim 0.2$).
As $\Phi_F$ reflects the gravitational potential, it can be measured even outside the Sun.
However, $A_F$, which reflects the mass density, falls steeply near the surface.
Figure \ref{fig_ampf-of-y} is consistent with Figs.~6 and 7 of \cite{Marchant2020}.

\begin{figure}
\epsscale{1.25}
\plotone{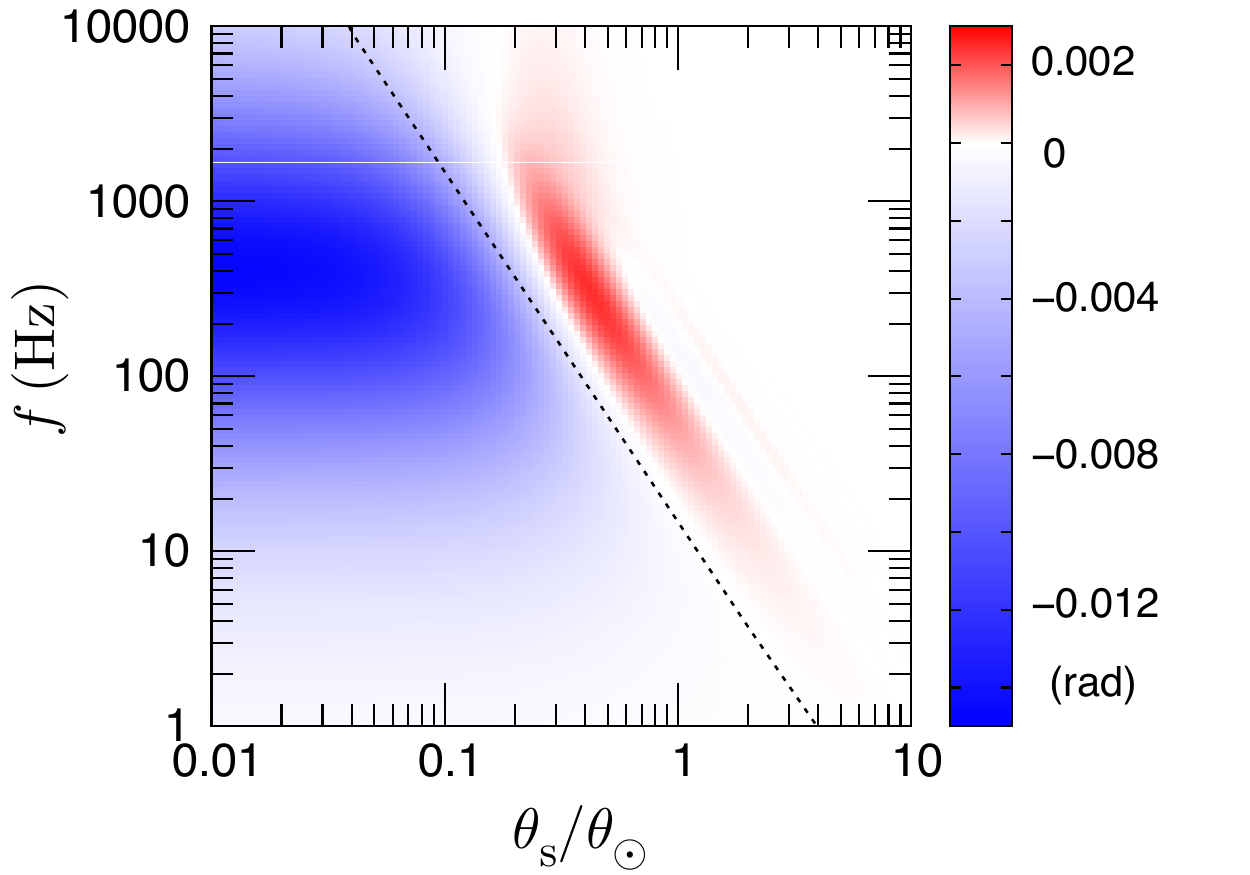}    
\caption{Contour plot of the difference between the phase modulation and its geometrical-optics limit ($\Phi_F-2 \pi f t_{\rm d}$). Red, blue, and white regions correspond to positive, negative, and zero differences, respectively. Dotted line plots the rough boundary between the wave- and geometrical-optics regions, calculated from $\theta_{\rm F}=\theta_{\rm s}$ (the description after Eq.~(\ref{Fresnel}) in the text).} 
\label{fig_dPhi}
\vspace*{0.3cm}
\end{figure}

The LVK analysis accounts for the Shapiro time delay imposed by the Sun. 
In searches of continuous GWs~\citep{Abbott2017,Abbott2019,Abbott2022}, the time delay has been calculated using the {\tt TEMPO}~\citep{Nice2015} and {\tt TEMPO2}~\citep{Edwards2006,Hobbs2006} packages developed for timing analyses of pulsar radio signals. 
However, as shown in Figs.~\ref{fig_ampf} and \ref{fig_ampf-of-y}, the Shapiro time delay is less accurate at lower $f$ and smaller $\theta_{\rm s}$.
To clarify the accuracy of the Shapiro time delay, Fig. \ref{fig_dPhi} plots a contour map of the difference between the phase modulation and its geometrical-optics limit. 
The difference ranges from $-0.015$ to $0.002$ rad and is most significant at $f\simeq100$--$1000 \, {\rm Hz}$ near the center ($\theta_{\rm s} \lesssim 0.2 \, \theta_\odot$).
At lower frequencies ($f \lesssim 10 \, {\rm Hz}$), the difference reduces because both $\varPhi_F$ and $2 \pi f t_{\rm d}$ are proportional to $f$.
As expected, the Shapiro time delay is recovered at high $f$ or large $\theta_{\rm s}$. 
The dotted line, predicted by setting $\theta_{\rm F}=\theta_{\rm s}$, is apparently consistent with the boundary.
Although the Shapiro time delay is less accurate inside the Sun, the error is small ($0.015$ rad at most).



\section{Known pulsars}
\label{sec:pulsars}

\begin{deluxetable*}{lrrrccc}
\tablecaption{Pulsars with $f\!>\!10 \, {\rm Hz}$ crossing behind the Sun (upper four) and close to, but not behind, the Sun (lower eight)}
\tablehead{
\colhead{name} & \colhead{$f\,$(Hz)} & \colhead{$\theta_{\rm {\scriptscriptstyle LAT}}/\theta_\odot$ } & \colhead{$D_{\rm s}\,$(kpc)} & \colhead{$h_0^{95 \%}$} & \colhead{$\epsilon^{95 \%}$} 
& \colhead{references}
}
\startdata
J1745$-$23 & $369$~ & $-0.07$~ & $7.94$ & ------ & ------ 
& C20 \\
J1022+1001 & $122$~ & $-0.24$~ & $0.64$ & $7.7 \!\times\! 10^{-27}$ & $3.2 \!\times\! 10^{-7}$ 
& C96,R21 \\
J1809$-$2332 & $14$~ & $-0.45$~ & $0.88$ & ------ & ------ 
& A09,R11 \\
J1730$-$2304 & $246$~ & $0.71$~ & $0.47$ & $5.0 \!\times\! 10^{-27}$ & $3.7 \!\times\! 10^{-8}$ &  L95,R21 \\
\hline
J1858-2216 & $839$~ & $1.80$~ & $0.92$ & $7.8 \!\times\! 10^{-27}$ & $9.6 \!\times\! 10^{-9}$ & S16 \\
J1142+0119 & $394$~ & $-1.84$~ & $2.17$ & $7.4 \!\times\! 10^{-27}$ & $9.8 \!\times\! 10^{-8}$ & S16 \\
J2310-0555 & $766$~ & $-1.99$~ & $1.56$ & ------ & ------ & S16 \\
J1756-2251 & $70$~ & $2.14$~ & $0.73$ & $6.1 \!\times\! 10^{-27}$ & $8.6 \!\times\! 10^{-7}$ & F04,F14 \\
J1646-2142 & $342$~ & $2.44$~ & $0.97$ & ------ & ------ & R12,R13 \\
J1811-2405 & $752$~ & $-2.52$~ & $1.83$ & $1.5 \!\times\! 10^{-26}$ & $4.5 \!\times\! 10^{-8}$ & K10,N20\\
J1836-2354B & $619$~ & $-2.74$~ & $3.20$ & ------ & ------ & L11 \\
J1836-2354A & $596$~ & $-2.78$~ & $3.20$ & ------ & ------ & L11 
\enddata
\tablecomments{Second column: GW frequency. Third column: ecliptic latitude $\theta_{\rm {\scriptscriptstyle LAT}}$ (i.e., the minimum angular separation to the solar center). Pulsars are listed in order of increasing $|\theta_{\rm {\scriptscriptstyle LAT}}|$. Fourth column: distance $D_{\rm s}$, measured by parallax for J1022+1001, J1730-2304, and J1756-2251 and estimated from the dispersion measures with the Galactic free-electron distribution model YMW16~\citep{YMW2017} for the other pulsars. Fifth and sixth columns: $95 \%$ upper limits on the strain amplitude and ellipticity, respectively, taken from the LVK results~\citep{Abbott2022}. Last column: references A09~\citep{Abdo2009}, C20~\citep{Cameron2020}, C96~\citep{Camilo1996}, F04~\citep{Faulkner2004}, F14~\citep{Ferdman2014}, K10~\citep{Keith2010}, L95~\citep{Lorimer1995}, L11~\citep{Lynch2011}, N20~\citep{Ng2020}, R11~\citep{Ray2011}, R12~\citep{Ray2012}, R21~\citep{Reardon2021}, R16~\citep{Roy2013}, and S16~\citep{Sanpa2016}.}
\label{table_pulsars}
\end{deluxetable*}

This section describes some known pulsars moving behind the Sun. 
From the results in Subsection \ref{sec:ap_modulation}, a pulsar with a higher frequency and a smaller impact parameter to the Sun is more suitable for probing the solar structure.
We searched the ATNF pulsar catalog\footnote{http://www.atnf.csiro.au/research/pulsar/psrcat/} for pulsars satisfying the following criteria: i) $f\!>\!10 \,$Hz and ii) closest distance to the solar center is within $3 \, \theta_\odot$.
Twelve samples met the criteria: four pulsars crossing behind the Sun and eight pulsars passing near but not behind the solar surface.
The information of these pulsars is summarized in Table \ref{table_pulsars}.
The GW frequency $f$ is twice the spin frequency for the GW emission from the mass-quadrupole moment.
These frequencies are covered by the sensitivity range of the current ground-based detectors.
The LVK collaboration obtained $95\%$ credible upper limits on the strain amplitudes of some of these samples from the O2 and O3 runs~\citep{Abbott2022}.
They also presented the corresponding upper bounds on the ellipticity of the mass distribution (estimated from Eq.~(\ref{h0_pulsar})).
The collaboration also performed an all-sky search for continuous GWs coming from any (known and unknown) spinning neutron stars in isolated and binary systems~\citep{Abbott2021,LVK2022}.
Their upper bound was $h_0^{95 \%} \sim 1.1 \, (2.0) \times 10^{-25}$ at $f=100$--$200$ $(50$--$700) \, {\rm Hz}$ for isolated systems (and similarly for binary systems).

\begin{figure}
\epsscale{1.1}
\plotone{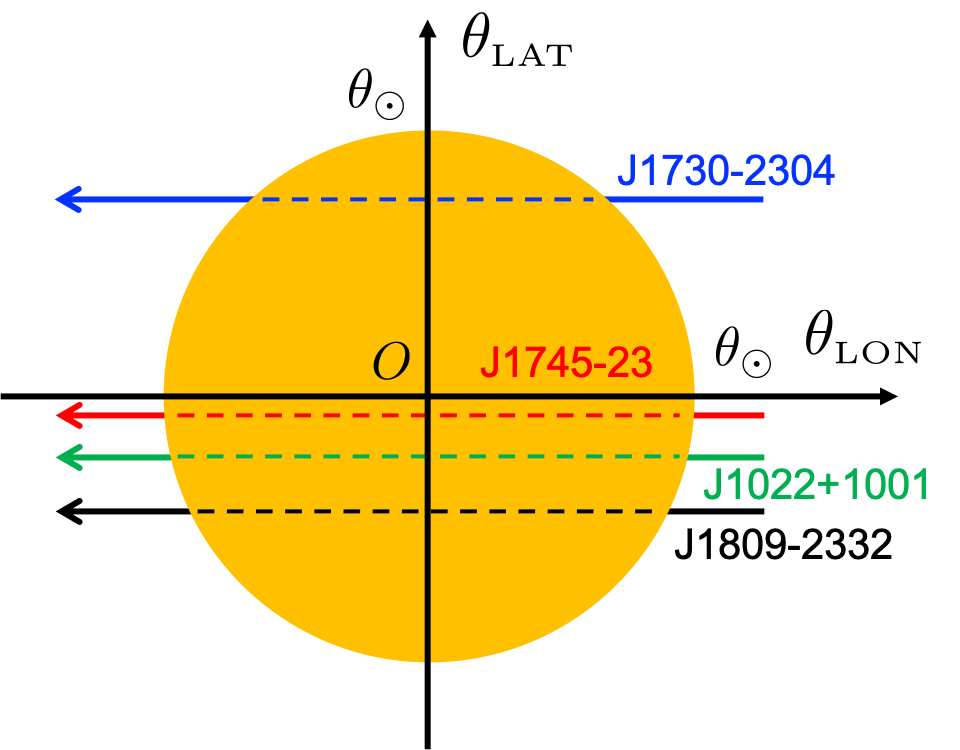}  
\caption{Trajectories of pulsars moving behind the Sun (represented by an orange circle of radius $\theta_\odot$). $\theta_{\rm {\scriptscriptstyle LAT}}$ is the ecliptic latitude and $\theta_{\rm {\scriptscriptstyle LON}}$ is the ecliptic longitude with origin set at the solar center.} 
\label{fig_pulsar_trajs}
\vspace*{0.3cm}
\end{figure}

Figure \ref{fig_pulsar_trajs} shows the trajectories of the upper four pulsars in Table \ref{table_pulsars}. 
These pulsars move along the horizontal axis with a velocity of $\dot{\theta}_{\rm {\scriptscriptstyle LON}} \simeq -2\pi/(1 {\rm yr})=-2.5 \, {\rm arcmin}/{\rm hour}=-0.15 \, \theta_\odot/{\rm hour}$. 
The typical crossing time is $2 \theta_\odot/|\dot{\theta}_{\rm LON}| \simeq 13\,{\rm hours}$. 
\cite{Marchant2020} included J1022+1001 and J1730-2304 as candidates, but excluded the very recently discovered pulsar J1745-23~\citep{Cameron2020}.  

\begin{figure*}
\vspace*{0cm}
\begin{minipage}{\columnwidth}
\includegraphics[width=1.2\columnwidth]{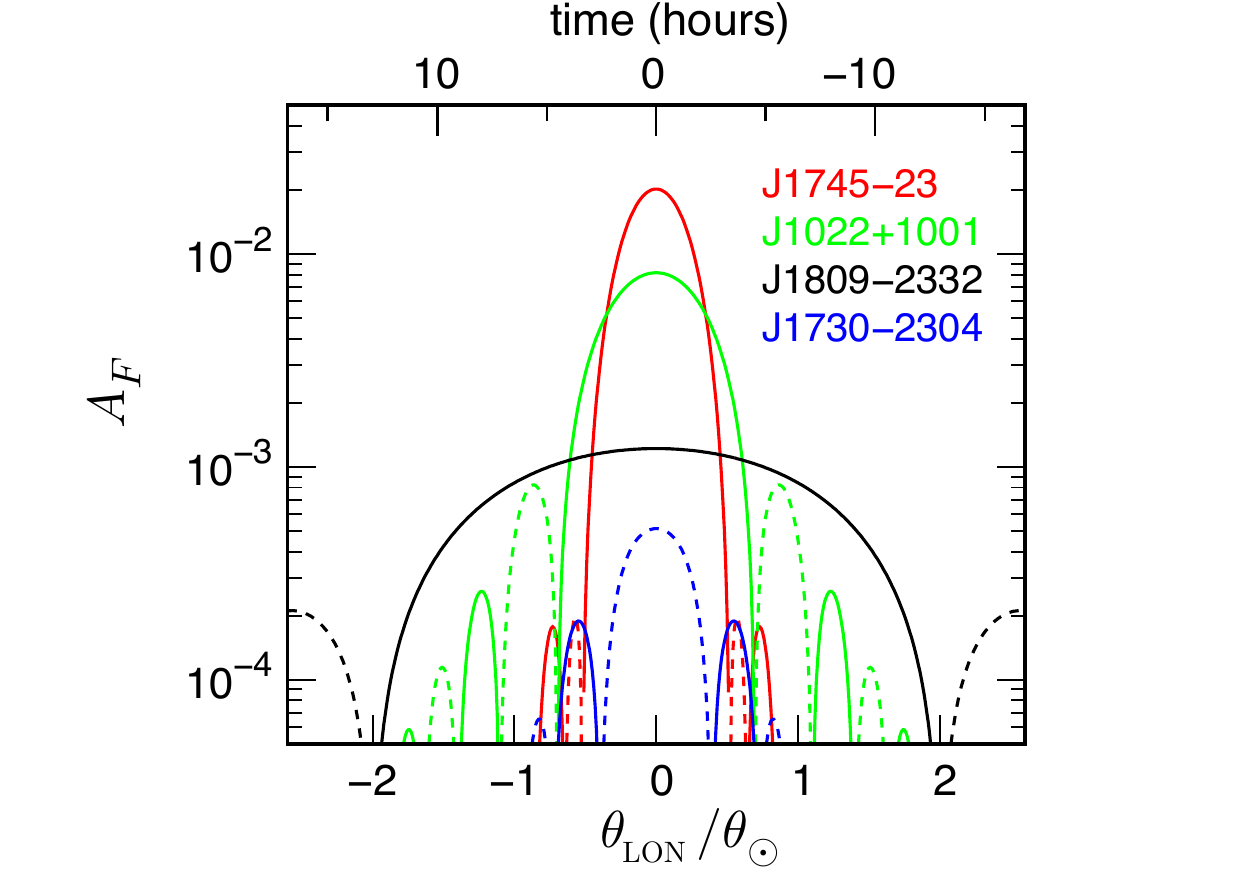}   
\end{minipage}
\begin{minipage}{\columnwidth}
\includegraphics[width=1.2\columnwidth]{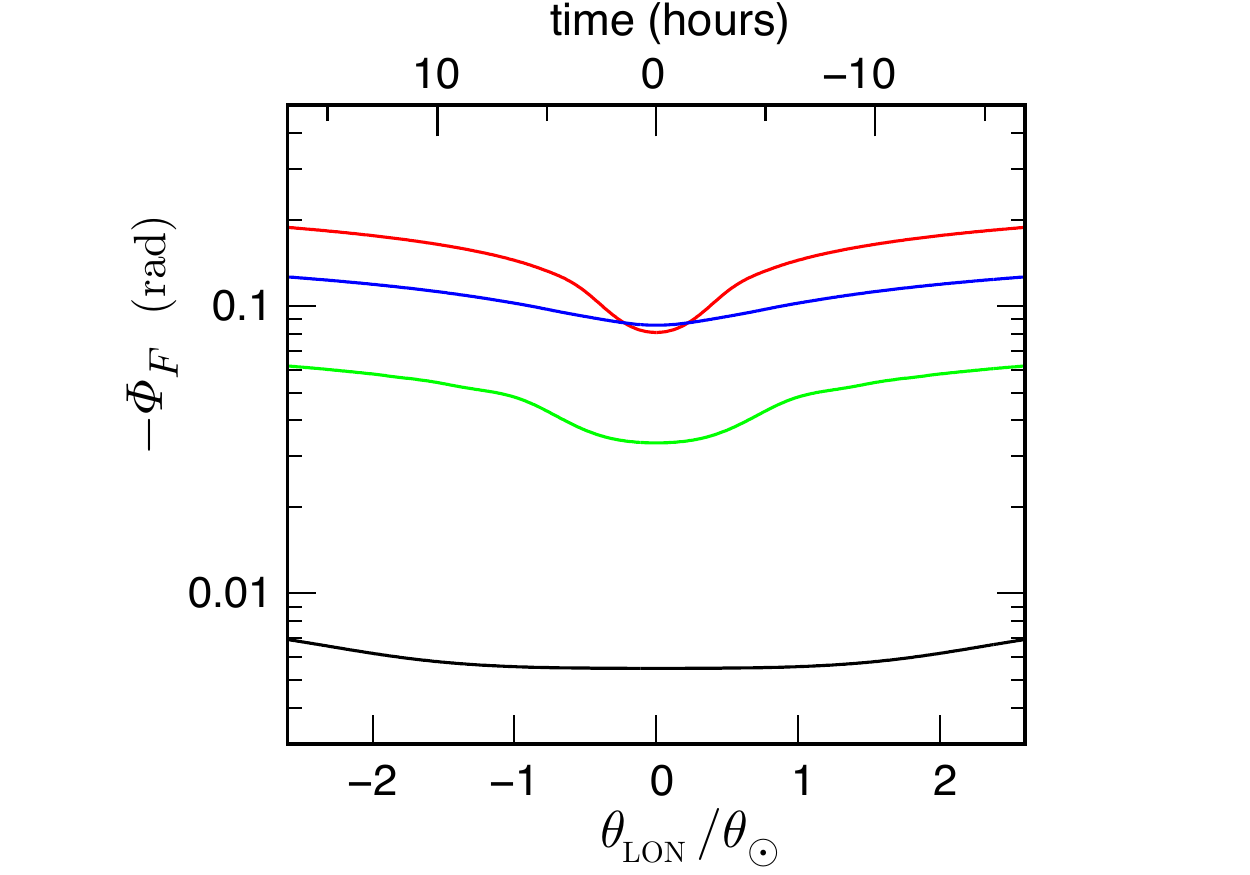}  
\end{minipage}
\caption{Amplitude and phase modulations of the four pulsars moving behind the Sun. The lower $x$ axis is the pulsar position $\theta_{\rm {\scriptscriptstyle LON}}/\theta_\odot$ along the ecliptic longitude (Fig.~\ref{fig_pulsar_trajs}), while the upper $x$ axis represents the corresponding crossing time. Dotted curves in the left panel represent negative values.}
\vspace*{0.3cm}
\label{fig_ampf_pulsars}
\end{figure*}

Figure \ref{fig_ampf_pulsars} plots $A_F$ and $\varPhi_F$ as functions of $\theta_{\rm {\scriptscriptstyle LON}}$ for the four pulsars.
J1745-23, with the highest frequency and closest impact parameter, yields the maximum $A_F$ ($\sim 2 \, \%$).
The highest $|\varPhi_F|$ is $\sim 0.1\,$rad for all pulsars except J1809-2332.
The anomalous result for J1809-2332 is attributable to the low frequency ($14\,$Hz) of this pulsar.
The typical $|\varPhi_F|$ is estimated as $|\varPhi_F| \simeq 2 \pi f |t_{\rm d}|$ $\simeq 0.1 \, {\rm rad} \, (f/200 \, {\rm Hz}) \, (|t_{\rm d}|/ 100 \, \mu {\rm s})$. 
The right panel clearly shows the potential well around $\theta_{\rm {\scriptscriptstyle LON}}=0$.
The effect is especially noticeable in the curves of J1745-23 and J1022+1001. 
As evidenced in the figure, $|A_F| \ll |\varPhi_F|$. 

\begin{figure}
\epsscale{1.4}
\plotone{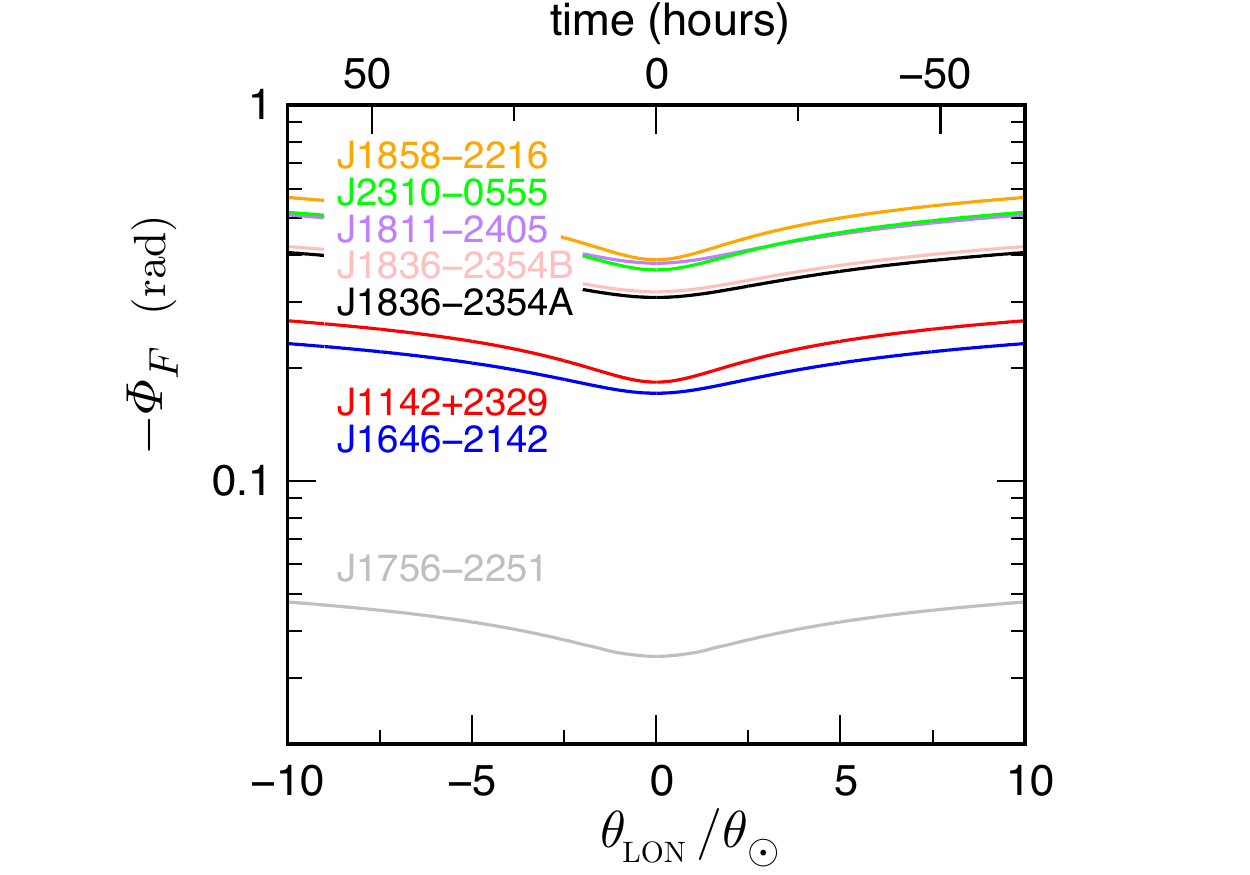}   
\caption{Phase modulations of the eight pulsars passing close to (but not behind) the Sun.} 
\label{fig_ampf_pulsars2}
\vspace*{0.3cm}
\end{figure}

Figure \ref{fig_ampf_pulsars2} plots $\varPhi_F$ as a function of $\theta_{\rm {\scriptscriptstyle LON}}$ for the lower eight pulsars in Table \ref{table_pulsars}.
Because these pulsars (except for J1756-2251) have relatively high frequencies, their $|\varPhi_F|$ are relatively large ($>0.1$ rad).  
As these pulsars do not pass behind the Sun, their $|A_F|$ are negligibly small.

\section{Parameter extraction from the lensed waveform}
\label{sec:fisher}

This section evaluates the accuracy to which the solar density profile can be extracted from the lensed GW signal through a Fisher analysis.

\subsection{Fisher matrix analysis}

We consider a highly spinning neutron star with a small non-axisymmetry around the spin axis as the GW source.
Neglecting spin down, the source is assumed to emit continuous monochromatic GWs during the observational period.
A waveform of frequency $f$ is then described as 
\beq
  h(t;f)=h_0 \cos \left( 2 \pi f t + \phi_0 \right),
  \label{unlensed_waveform}
\eeq
where $\phi_0$ is a constant phase. 
The strain amplitude $h_0$ is determined by the non-axisymmetry as~\citep[e.g.,][]{Riles2022}
\beq
   h_0=\frac{4 \pi^2 \epsilon G I f^2}{c^4 D_{\rm s}},
   \label{h0_pulsar}
\eeq
where $I$ is the moment of inertia about the spin axis and $\epsilon$ is the ellipticity of the moment of inertia.
A gravitationally lensed waveform $h^{\rm L}(t;f)$ can be obtained from Eqs.~(\ref{AF_PhiF}) and (\ref{unlensed_waveform}) as\footnote{Here, we perform a Fourier transform to obtain the unlensed waveform in the frequency domain, then calculate the lensed waveform using Eq.~(\ref{AF_PhiF}), and finally repeat the Fourier transform to obtain the lensed waveform (\ref{lensed_waveform}) in the time domain.}
\beq
 h^{\rm L}(t;f) = h_0 \left[ 1+A_F(t;f) \right] \cos \left[ 2 \pi f t + \phi_0 + \varPhi_F(t;f) \right].
\label{lensed_waveform}
\eeq
Here, $A_F$ and $\varPhi_F$ are numerically obtained from Eqs.~(\ref{AF_PhiF}) and (\ref{ampf2}).
When $\theta_{\rm s}$ is large ($\theta_{\rm s} > 20 \, \theta_\odot$), we can apply the geometrical-optics results (Eqs.~(\ref{AP_geo}) and (\ref{psi_largetheta})), which are fully valid at such large angles (Fig.~\ref{fig_ampf-of-y}).


Suppose that the pulsar is observed from time $t=-T/2$ to $T/2$, where $T$ is the observational period and $t=0$ is the time at which the pulsar is closest to the solar center.
The signal-to-noise ratio (SNR) is calculated as
\begin{align}
    {\rm SNR}^2 &= \frac{1}{S_n(f)} \int_{-T/2}^{T/2} \! dt \left[ h^{\rm L}(t;f) \right]^2, \nonumber \\
    & \simeq \frac{h_0^2}{2 S_n(f)} \, T,
\label{sn}
\end{align}
where $S_n$ is the noise spectrum of the detector.
Throughout this paper, the measurement accuracies of the fitting parameters are normalized by the SNR. Therefore, the results do not depend on a specific form of $S_n$.
In the second equality of Eq.~(\ref{sn}), we neglect $A_F$ and use the approximation $\cos^2 (2 \pi f t + \phi_0 + \varPhi_F) \simeq 1/2$ obtained by long-time averaging.
We comment on the validity of the approximation in Eq.~(\ref{sn}). 
The left panel of Fig.~\ref{fig_ampf_pulsars} shows $A_F \approx \mathcal{O}(0.01)$ for a few hours just behind the Sun; therefore, its average over one year is $\langle A_F \rangle_{\rm 1yr} \approx \mathcal{O}(0.01) \times ({\rm a~few~hours})/(1 {\rm yr}) \approx \mathcal{O}(10^{-6})$. Similarly, in the phase of Eq. (\ref{lensed_waveform}), the first term, $2 \pi ft \approx 6 \times 10^{10} (f/{\rm 300 \, Hz}) (t/1 \, {\rm yr})$, is ten orders of magnitude larger than the third term, $\varPhi_F \approx \mathcal{O}(1)$ rad from Figs. 7 and 8. This suggests that $A_F$ and $\varPhi_F$ give negligible contributions to the SNR (but note that $A_F$ does increase the SNR during the crossing).
This also suggests that $h_L(t;f)$ is rapidly oscillating during the observation and thus the approximation $\cos^2 (2 \pi f t + \phi_0 + \varPhi_F) \simeq 1/2$ is valid with a high precision.

The lensed waveform (\ref{lensed_waveform}) depends on the solar density profile, the constant amplitude $h_0$, and the constant phase $\phi_0$. 
Let the solar density profile be characterized by $N$ parameters $A_{\rm i}$ (${\rm i}=1,2,\cdots,N$).
The total set of fitting parameters is $p_\mu=(A_{\rm i},\ln h_0,\phi_0)$ with $\mu=1,2,\cdots,N+2$. 
Let us consider a GW signal as a sum of the theoretical template (\ref{lensed_waveform}) characterized by the fiducial parameters $\bar{p}_\mu$ and a Gaussian noise.
Here, our fiducial solar model is the BS05(OP) in \cite{Bahcall:2005}.
In a likelihood analysis, one can obtain the best-fit parameters by fitting the theoretical template to the signal.
In the limit of high SNR, the posterior distribution of ${\bm p}$ is multivariate Gaussian: $P({\bm p}) \propto \exp [- \sum_{\mu,\nu} (p_\mu-\bar{p}_\mu) \Gamma_{\mu\nu} (p_\nu-\bar{p}_\nu)/2 \, ]$~\citep[for a detailed discussion, see][]{CF1994}.
The measurement accuracy of $p_\mu$ is given by the inverse of the Fisher matrix: $\Delta p_\mu = [ (\Gamma^{-1})_{\mu\mu} ]^{1/2}$.
Neglecting the degeneracy with the other parameters, $\Delta p_\mu = (\Gamma_{\mu\mu})^{-1/2}$.
The Fisher matrix is given by~\citep[e.g.,][]{Cutler1998} 
\begin{equation}
    \Gamma_{\mu\nu} = \frac{1}{S_n(f)} \int_{-T/2}^{T/2} \! dt \, \frac{\partial h^{\rm L}(t;f)}{\partial p_\mu} \frac{\partial h^{\rm L}(t;f)}{\partial p_\nu}.
\end{equation}
Using Eqs.~(\ref{lensed_waveform}) and (\ref{sn}), all components of $\Gamma$ are given as follows:
\begin{align}
    \Gamma_{A_{\rm i} A_{\rm j}} &= \int_t \, \left[ \frac{\partial A_F(t;f)}{\partial A_{\rm i}} \frac{\partial A_F (t;f)}{\partial A_{\rm j}} \right.  \nonumber \\
   &~~ + \left\{ 1+A_F(t;f) \right\}^2 \left. \frac{\partial \varPhi_F(t;f)}{\partial A_{\rm i}} \frac{\partial \varPhi_F (t;f)}{\partial A_{\rm j}} \right],  \nonumber \\
   \Gamma_{A_{\rm i} \ln \! h_0} &= \int_t \,  \left\{ 1+A_F(t;f) \right\} \frac{\partial A_F(t;f)}{\partial A_{\rm i}},  \nonumber \\
   \Gamma_{A_{\rm i} \phi_0} &= \int_t \, \left\{ 1+A_F(t;f) \right\}^2 \frac{\partial \varPhi_F(t;f)}{\partial A_{\rm i}},  \nonumber \\
   \Gamma_{\ln \! h_0 \ln \! h_0} &= \Gamma_{\phi_0 \phi_0} = \int_t \, \left\{ 1+A_F(t;f) \right\}^2,  \nonumber \\
   \Gamma_{\ln \! h_0 \phi_0} &= 0,  
   \label{fisher}
\end{align}
with
\begin{equation}
     \int_t \, \equiv {\rm SNR}^2 \, \frac{1}{T} \int_{-T/2}^{T/2} \! dt. \nonumber
\end{equation}
Similar to Eq.~(\ref{sn}), we here applied the approximations $\sin^2 {\rm or} \cos^2 (2 \pi f t + \phi_0 + \varPhi_F) \simeq 1/2$ and $\sin (2 \pi f t + \phi_0 + \varPhi_F) \, \cos (2 \pi f t + \phi_0 + \varPhi_F) \simeq 0$. 
From Eq.~(\ref{fisher}), $\Delta p_\mu$ simply scales as $\Delta p_\mu \propto {\rm SNR}^{-1}$.

The simple waveform (\ref{unlensed_waveform}) neglects the detector response which depends on other source parameters such as the inclination of the spin axis and GW polarization. 
Owing to Earth's spin and orbital motion, the detector response causes daily and yearly periodic modulations on both amplitude and phase.
Unlike the solar modulation (which is temporary but repeats annually), the response modulations are trigonometric functions of time. 
We expect that the solar modulation can be distinguishable from the response modulations and thus the response does not significantly affect the measurement accuracies of the solar model parameters. 


\subsection{Constraint on the potential amplitude}
\label{subsec:fisher-potential}

\begin{figure*}
\epsscale{1.1}
\vspace{-4cm}
\plotone{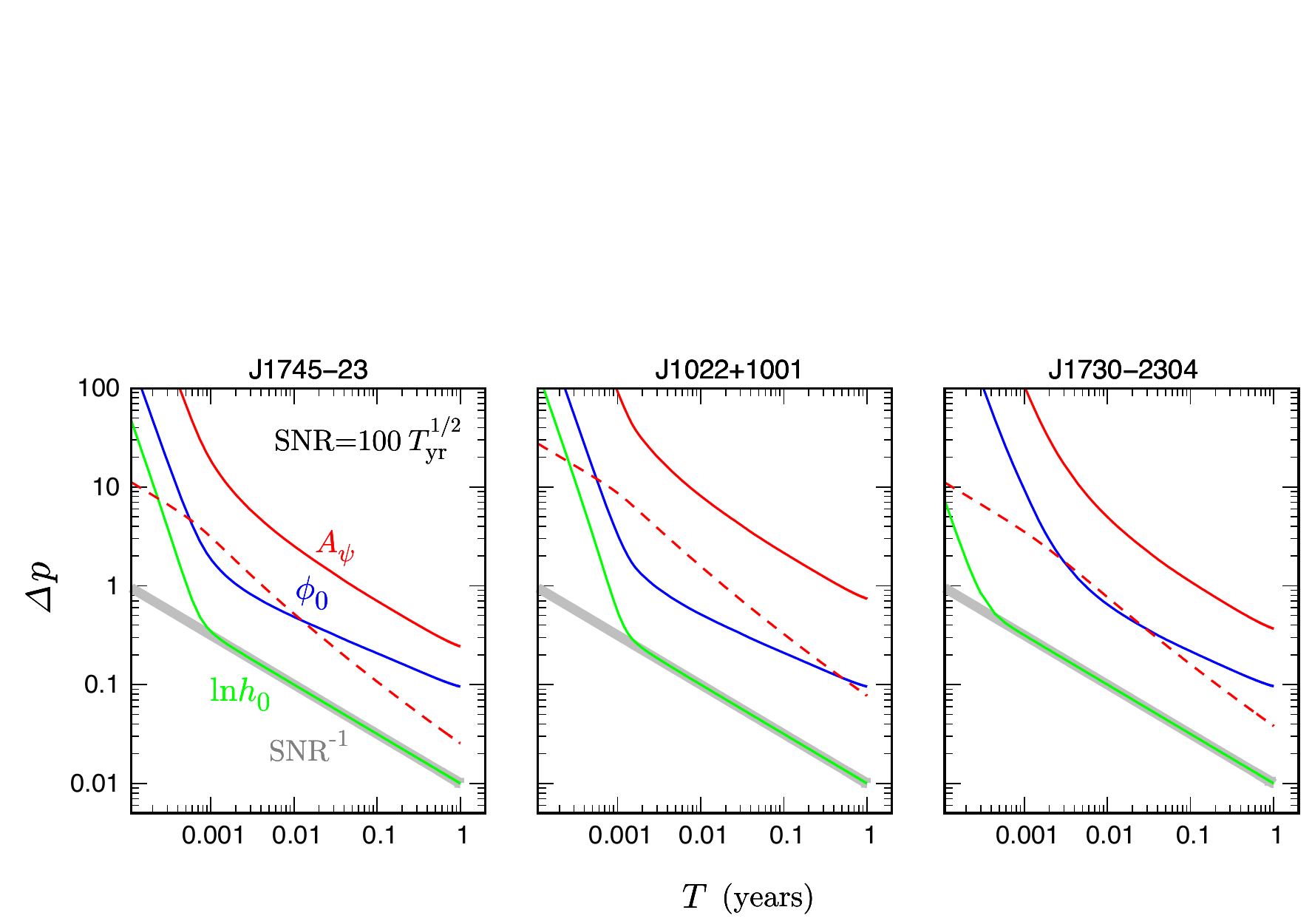}  
\caption{Measurement accuracies of the solar-potential amplitude $A_\psi$ defined in Eq.~(\ref{A_psi}), amplitude $h_0$, and phase $\phi_0$ of the pulsars crossing behind the Sun. 
Dashed red curve plots $\Delta A_\psi$ neglecting the parameter degeneracy. The horizontal axis $T$ denotes the observational period. The signal-to-noise ratio (SNR) is set to $100 \, (T/{\rm yr})^{1/2}$ and the results simply scale as $\Delta p \propto {\rm SNR}^{-1}$. Thick gray line denotes ${\rm SNR}^{-1}$.} 
\label{fig_Am_error_pulsar}
\vspace*{0.3cm}
\end{figure*}

This subsection presents the measurement accuracy of the overall amplitude of the solar gravitational potential.
Replacing the potential as
\beq
  \hat{\psi}(\theta) \rightarrow A_\psi \hat{\psi}(\theta),
  \label{A_psi}
\eeq
we calculate the measurement error in $A_\psi$.
The derivatives in the Fisher matrix with respect to $A_\psi$ are numerically obtained by changing $A_\psi$ by $\pm 1  \, \%$. 

Figure \ref{fig_Am_error_pulsar} plots the measurement accuracies of $A_\psi$, $\ln h_0$ and $\phi_0$ for the pulsars moving behind the Sun at $f>100 \, {\rm Hz}$.
The results are normalized with ${\rm SNR}=100$ over a one-year observation period. 
Because $|A_F| \ll |\Phi_F|$, $A_\psi$ is mainly determined by the phase modulation.
The slope of $\Delta A_\psi$ changes slightly at the surface (corresponding to $T \approx 10^{-3} \, {\rm yr}$); the slope is steeper (shallower) inside (outside) the Sun. 
These results can be explained as follows. 
In the Sun's interior, $|\varPhi_F|$ increases with $\theta$ (in contrast to the constant $\phi_0$), thus $A_\psi$ and $\phi_0$ can be determined almost independently. 
Outside the Sun, $\varPhi_F$ is nearly constant at large $\theta \, (\gg \theta_\odot)$, thus $A_\psi$ and $\phi_0$ are degenerate to some extent.
The dashed red curve plots $\Delta A_\psi$ without the parameter degeneracy (i.e., $\Delta A_\psi=(\Gamma_{A_\psi A_\psi})^{-1/2}$), which is $\sim 10$ times better than the solid red curve.
Previously, \cite{Marchant2020} roughly estimated the detectability of lensing signatures.
As they ignored the parameter degeneracy, they underestimated the measurement error. 
Inside the Sun, $h_0$ cannot be determined, because $h_0$ and $A_F$ are highly degenerate, but outside the Sun, the degeneracy is broken and $\Delta \ln h_0 \simeq {\rm SNR}^{-1}$.


We caution that $\Delta \phi_0$ shown in Fig.~{\ref{fig_Am_error_pulsar}} depends on an arbitrary constant in the gravitational potential.
If a constant term is added to the potential, i.e., $\hat{\psi}(\theta) \rightarrow \hat{\psi}(\theta)+\hat{\psi}_0$, the phase modulation changes as $\varPhi_F \rightarrow \varPhi_F- 2 \pi f \hat{\psi}_0/c^3$ using Eqs.~(\ref{AF_PhiF})--(\ref{time_delay}).
Two components of the Fisher matrix change accordingly\footnote{$\Gamma_{A_\psi \phi_0} \rightarrow \Gamma_{A_\psi \phi_0} - (2 \pi f \hat{\psi}_0/c^3) \Gamma_{\phi_0 \phi_0}$ and $\Gamma_{A_\psi A_\psi} \rightarrow \Gamma_{A_\psi A_\psi} - (4 \pi f \hat{\psi_0}/c^3) \Gamma_{A_\psi \phi_0} + (2 \pi f \hat{\psi}_0/c^3)^2 \Gamma_{\phi_0 \phi_0}$.}.
After some algebra, one finds that $\Delta A_\psi$, $\Delta \ln h_0$, and $(\Gamma^{-1})_{A_\psi \ln h_0}$ are independent of $\hat{\psi}_0$ but $\Delta \phi_0$ and its cross correlations, $(\Gamma^{-1})_{A_\psi \phi_0}$ and $(\Gamma^{-1})_{\ln h_0 \phi_0}$, do depend on $\hat{\psi}_0$.

\begin{deluxetable}{lcc}
\tablecaption{Measurement accuracies of the potential amplitude $A_\psi$ (the fiducial value $=1$) over one year of observation with ${\rm SNR}=100$.}
\tablehead{
\colhead{name}~~ & \colhead{$\Delta A_\psi$} & \colhead{$f\Delta A_\psi$ (Hz)} 
}
\startdata
J1745$-$23 & $0.24$~ ($0.026$) & $90$ \\ 
J1022+1001 & $0.74$~ ($0.077$) & $90$ \\ 
J1809$-$2332 & $6.6$~ ($0.69$) & $90$ \\ 
J1730$-$2304 & $0.37$~ ($0.038$) & $91$ \\ 
\hline
J1858-2216 & $0.11$~ ($0.011$) & $92$ \\ 
J1142+0119 & $0.23$~ ($0.024$) & $92$ \\ 
J2310-0555 & $0.12$~ ($0.012$) & $92$ \\ 
J1756-2251 & $1.3$~ ($0.13$)  & $92$ \\ 
J1646-2142 & $0.27$~ ($0.028$) & $93$ \\ 
J1811-2405 & $0.12$~ ($0.013$) & $93$ \\ 
J1836-2354B & $0.16$~ ($0.016$) & $93$ \\ 
J1836-2354A & $0.15$~ ($0.015$)  & $93$ 
\enddata
\tablecomments{Values in parentheses are calculated without the parameter degeneracy.}
\label{table_Apsi}
\end{deluxetable}

Table \ref{table_Apsi} lists the $\Delta A_\psi$ results of all pulsars. 
Although $\Delta A_\psi$ varies among the pulsars, $f \Delta A_\psi$ is almost constant; i.e., it is independent of the impact parameter to the Sun.
$\Delta A_\psi$ can then be approximately fitted as
\beq
  \Delta A_\psi \approx 0.3 \left( \frac{f}{300 \, {\rm Hz}} \right)^{-1} \left( \frac{\rm SNR}{100} \right)^{-1},
\label{deltaApsi}
\eeq
indicating that a higher-frequency pulsar is more promising for detecting $A_\psi$.
According to Eq.~(\ref{deltaApsi}), $A_\psi$ can be detected at the $3 \sigma$ confidence level when ${\rm SNR} \approx 100 \, (f/300 {\rm Hz})^{-1}$.



\subsection{Measurement of the solar density profile}
\label{subsec:fisher-density}

\begin{figure}
\epsscale{1.}
\plotone{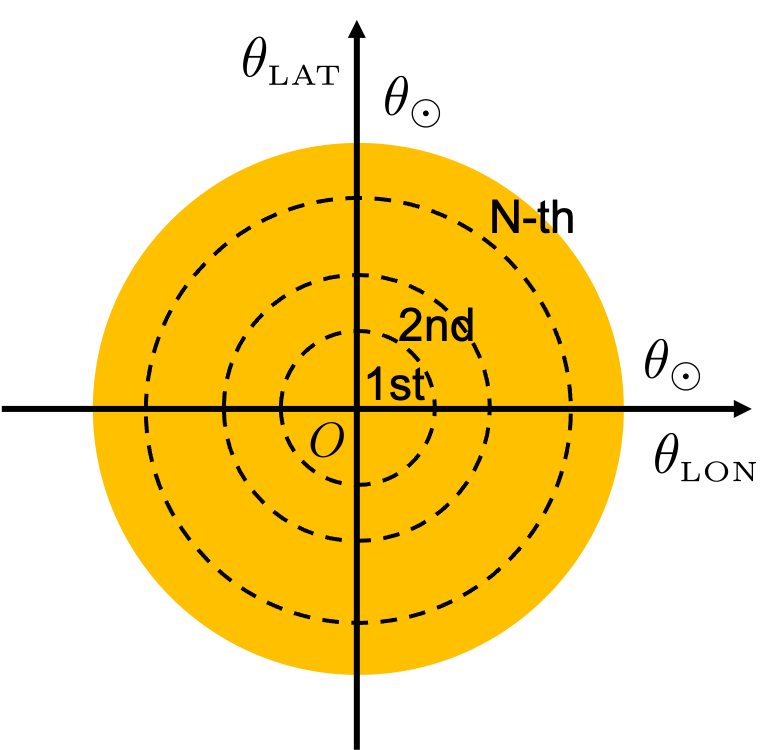}   
\caption{Division of the solar surface into $N$ annuli. 
Each annulus has the same mass in Case (I) and the same radial width in Case (II).} 
\label{fig_annuli}
\end{figure}

\begin{figure*}
\epsscale{1.1}
\vspace{-1cm}
\plotone{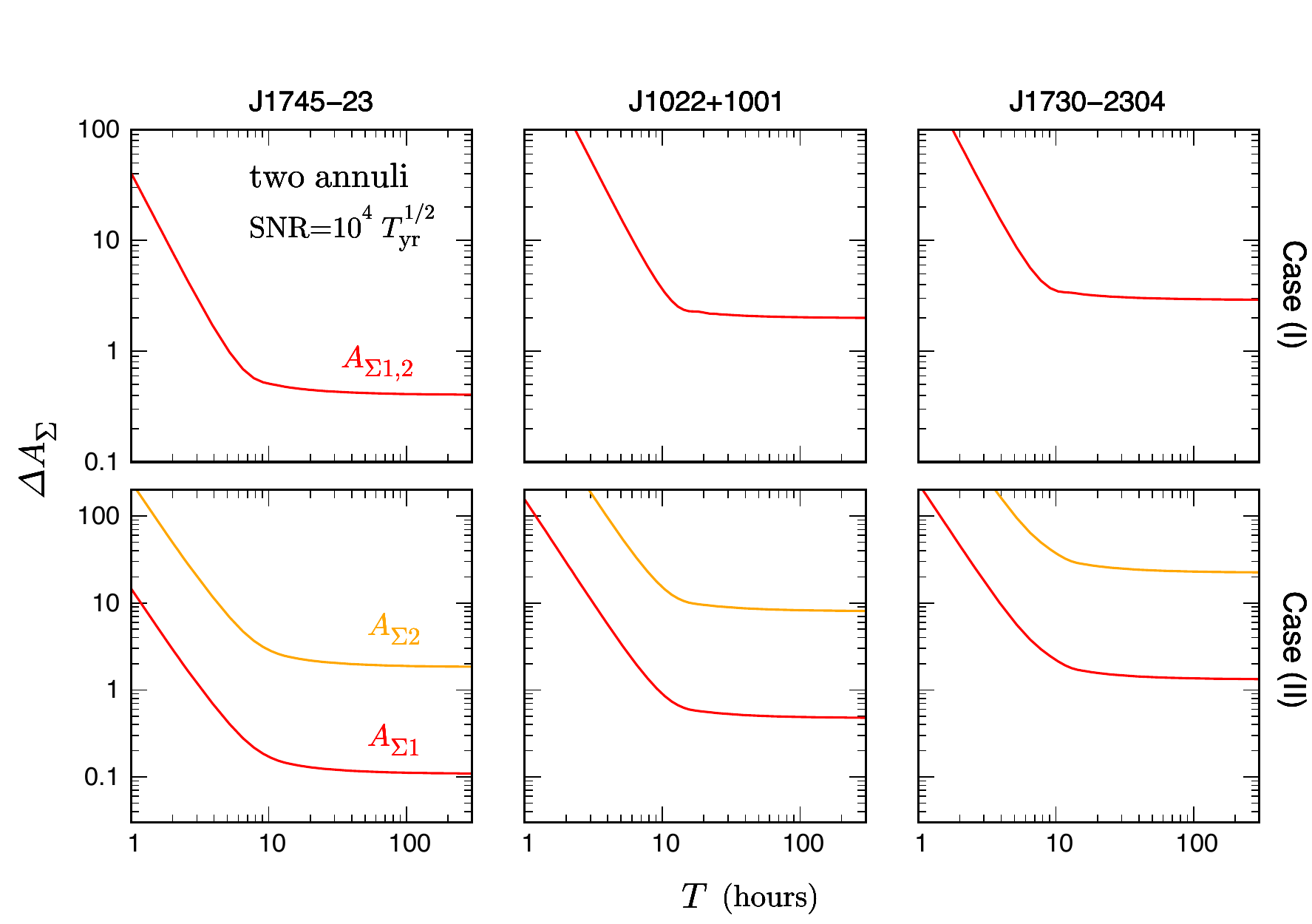}  
\caption{Measurement accuracies of the solar-density amplitudes in the i-th annulus $A_{\Sigma {\rm i}}$ for the pulsars crossing behind the Sun. Upper and lower panels plot the results in Case (I) and (II), respectively. In the upper panels, $\Delta A_{\Sigma 1}=\Delta A_{\Sigma 2}$. The horizontal axis $T$ denotes the observational period. SNR is set to $10^4 \, (T/{\rm yr})^{1/2}$ and the results simply scale as $\Delta A_{\Sigma {\rm i}} \propto {\rm SNR}^{-1}$.} 
\label{fig_A1A2_error}
\vspace*{0.3cm}
\end{figure*}

\begin{figure*}
\epsscale{1.1}
\vspace{-1cm}
\plotone{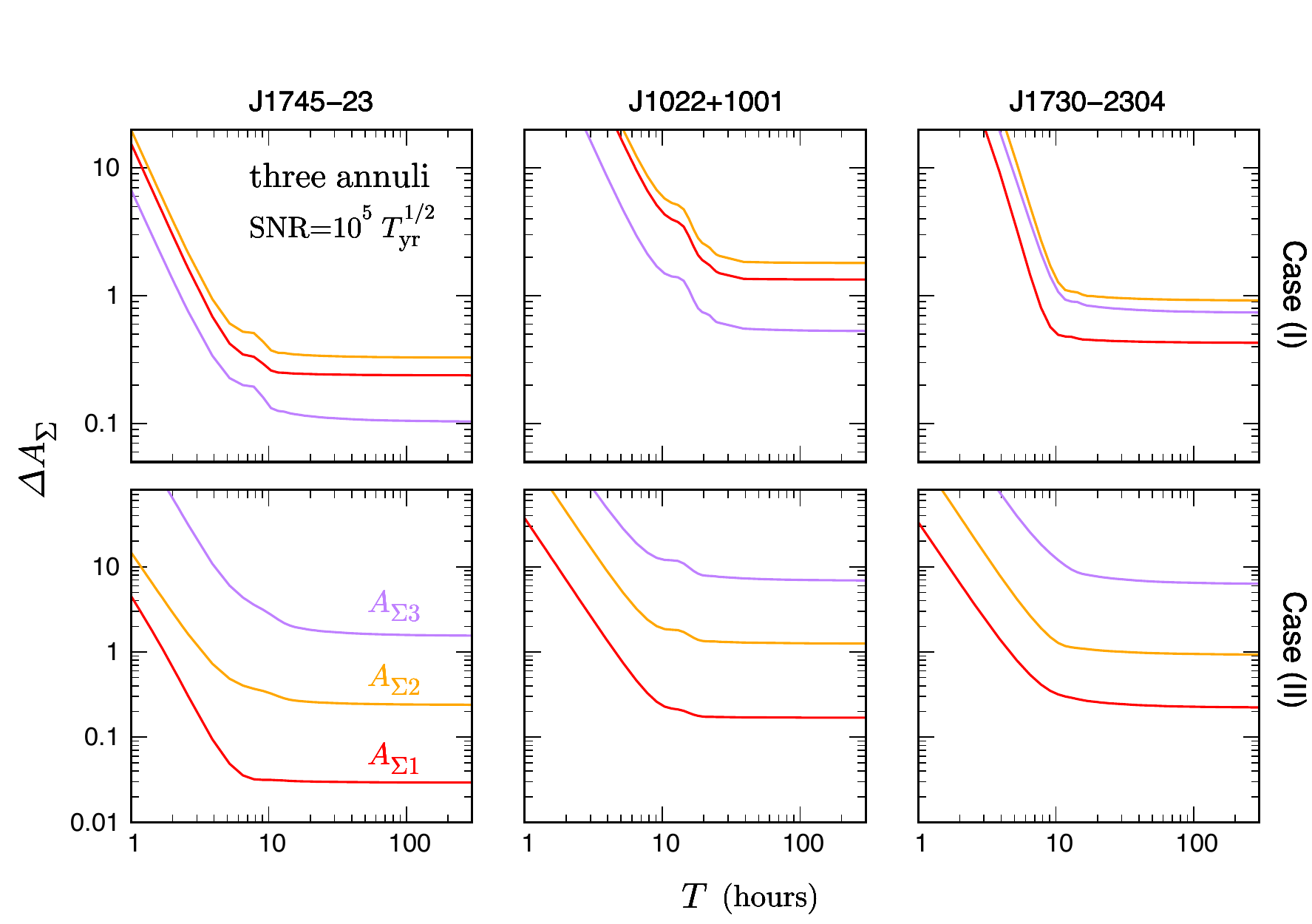} 
\caption{Same as Fig.~\ref{fig_A1A2_error}, but for three annuli with ${\rm SNR}=10^5 \, (T/{\rm yr})^{1/2}$.} 
\label{fig_A1A2A3_error}
\vspace*{0.5cm}
\end{figure*}

\begin{deluxetable}{lcc}
\tablecaption{Measurement accuracies of the density amplitudes in two annuli over a one-year observation period with ${\rm SNR}=10^4$.}
\tablehead{
 & \colhead{Case (I)} & \colhead{Case (II)} \\
\colhead{name}~~ & \colhead{$\Delta A_{\Sigma 1,2}$} & \colhead{$\Delta A_{\Sigma 1}, \, \Delta A_{\Sigma 2}$}}
\startdata
J1745$-$23 & $0.40$~ & $0.11,~1.8$ \\ 
J1022+1001 & $2.0$~ & $0.47,~8.0$ \\ 
J1730$-$2304 & $2.9$~ & $1.3,~22$
\enddata
\label{table_A1A2}
\tablecomments{In Case (I), $\Delta A_{\Sigma 1}=\Delta A_{\Sigma 2}$.}
\end{deluxetable}

\begin{deluxetable}{lcc}
\tablecaption{Same as Table \ref{table_A1A2}, but for three annuli and a one-year observation period with ${\rm SNR}=10^5$.}
\tablehead{
& \colhead{Case (I)} & \colhead{Case (II)} \\
\colhead{name}~~ & \colhead{$\Delta A_{\Sigma 1},\, \Delta A_{\Sigma 2},\, \Delta A_{\Sigma 3}$} & \colhead{$\Delta A_{\Sigma 1},\, \Delta A_{\Sigma 2},\, \Delta A_{\Sigma 3}$}}
\startdata
J1745$-$23 & $0.24,~0.33,~0.10$ & $0.029,~0.24,~1.5$ \\ 
J1022+1001 & $1.3,~1.8,~0.53$ & $0.17,~1.3,~6.9$ \\ 
J1730$-$2304 & $0.43,~0.92,~0.74$ & $0.22,~0.93,~6.3$ 
\enddata
\label{table_A1A2A3}
\end{deluxetable}

This subsection presents our main result, namely the measurement accuracy of the solar density profile.
Let the surface density be divided into $N$ annuli (Fig.~\ref{fig_annuli}). 
We write the surface mass density of the i-th annulus as $\Sigma_{\rm i}(\theta)$ and the mass as $M_{\rm i}$ (${\rm i}=1,\cdots,N$), where a smaller i corresponds to an inner annulus.
We consider the following two cases:
\begin{description}
   \item[Case (I)] Each annulus has the same mass ($=M_\odot/N$).
   \item[Case (II)] Each annulus has the same radial thickness ($=\theta_\odot/N$).
\end{description}
We first consider $N=2$ and $3$.
In Case (I), the boundary radius between the annuli is $0.20 \, \theta_\odot$ for $N=2$.
The boundary radii are $0.14 \, \theta_\odot$ and $0.26 \, \theta_\odot$ between annulus 1 and 2 and between annulus 2 and 3, respectively, for $N=3$.
In Case (II), the mass ratios are $M_1:M_2 \simeq 17:1$ for $N=2$ and $M_1:M_2:M_3 \simeq 57:13:1$ for $N=3$ (i.e., the mass reduces from the innermost to the outermost annulus).

We change the amplitude of the i-th density as
\beq
   \Sigma_{\rm i}(\theta) \rightarrow A_{\Sigma {\rm i}} \Sigma_{\rm i}(\theta),
\eeq
while fixing the total mass, i.e.,
 $\sum_{{\rm i}=1}^N A_{\Sigma {\rm i}} M_{\rm i} = M_\odot$.
Under these settings, we calculate the measurement accuracies of $A_{\Sigma {\rm i}}$, $\ln h_0$, and $\phi_0$. 
When $N=2$, the density profile is solely characterized by the fitting parameter $A_{\Sigma 1}$ because the total mass is fixed, i.e., $M_1 \Delta A_{\Sigma 1}=M_2 \Delta A_{\Sigma 2}$.
Similarly, when $N=3$, the fitting parameters are $A_{\Sigma 1}$ and $A_{\Sigma 2}$.
The other parameter, $A_{\Sigma 3}$, is determined via $(M_3 \Delta A_{\Sigma 3})^2=(M_1 \Delta A_{\Sigma 1})^2+(M_2 \Delta A_{\Sigma 2})^2+ 2 M_1 M_2  (\Gamma^{-1})_{A_{\Sigma 1} A_{\Sigma 2}}$. 
To obtain the derivative with respect to $A_{\Sigma {\rm i}}$ in the Fisher matrix, we numerically calculate the derivatives of the potential in Eq.~(\ref{Poisson}) and the amplification factor in Eq.~(\ref{ampf2}) by changing $A_{\Sigma {\rm i}}$ by $\pm 1 \, \%$. 

Figure \ref{fig_A1A2_error} plots the measurement accuracies of $A_{\Sigma {\rm i}}$ for $N=2$. 
As the total mass is fixed, the lensed signal is insensitive to the density profile when the pulsar is outside the Sun.
Therefore, the accuracies are not improved at $T \gtrsim 10 \, {\rm hours}$. 
J1745-23 gives the best accuracy, because it has the highest frequency and smallest impact parameter.
J1730-2304 has a higher frequency than J1022+1001, but its larger impact parameter lessens the constraint.
In Case (II), $\Delta A_{\Sigma 2}$ is $17$ times larger than $\Delta A_{\Sigma 1}$, reflecting the mass ratio $M_1/M_2 \simeq 17$.
According to the figure, at least ${\rm SNR} \approx 10^4 \, (T/{\rm yr})^{1/2}$ is required for probing the solar density profile, but a lower SNR may be sufficient for probing with J1745-23.
We comment that $\Delta \ln h_0$ and $\Delta \phi_0$ agree with ${\rm SNR}^{-1}$ outside the Sun (corresponding to $T>10$ hours; these results are not plotted in the figure).
The numerical values of $\Delta A_{\Sigma {\rm i}}$ are listed in Table \ref{table_A1A2}.

Figure \ref{fig_A1A2A3_error} plots the measurement accuracies of $A_{\Sigma {\rm i}}$ for $N=3$. 
In Case (I), $\Delta A_{\Sigma 3}$ shows the highest accuracy for most pulsars (the exception is J1730-2304), because the third annulus (with radius ranging from $0.26 \, \theta_\odot$ to $\theta_\odot$) has the largest area. 
Therefore, the pulsar remains longer in the third annulus than in the other annuli.
In Case (II), $\Delta A_{\Sigma 1} < \Delta A_{\Sigma 2} < \Delta A_{\Sigma 3}$ simply because the inner annulus is more massive than the outer one.
The numerical values of $\Delta A_{\Sigma {\rm i}}$ are listed in Table \ref{table_A1A2A3}.
According to Fig.~\ref{fig_A1A2A3_error}, at least ${\rm SNR} \approx 10^5 \, (T/{\rm yr})^{1/2}$ is required for measuring the density profile with $N=3$.
The SNR must be further increased for larger $N$ ($>3$).

In geometrical optics, $\varPhi_F$ is fully determined by the mass enclosed within a given radius.
Therefore, a pulsar can probe the enclosed mass down to the impact parameter. 
In wave optics, $\varPhi_F$ also depends on the outer density profile, owing to diffraction (Subsection \ref{sec:ap_modulation}).
Therefore, the accuracy of $A_{\Sigma {\rm i}}$ depends on the frequency and impact parameter in a complicated manner.
Combining the results of different pulsars improves the measurement accuracy of the solar density and increases the radial range of the estimation. 

Figures \ref{fig_A1A2_error} and \ref{fig_A1A2A3_error} suggest that the accuracies strongly depend on the chosen boundary.
In Case (I), the accuracies are similar in different annuli, but in Case (II), the accuracy is better in the inner annulus. 
This implies that the annulus mass mainly determines the accuracy. 
The boundary that maximizes the accuracy will depend on the frequency and impact parameter of the pulsar. 

\begin{figure*}
\vspace*{0cm}
\begin{minipage}{\columnwidth}
\includegraphics[width=1.2\columnwidth]{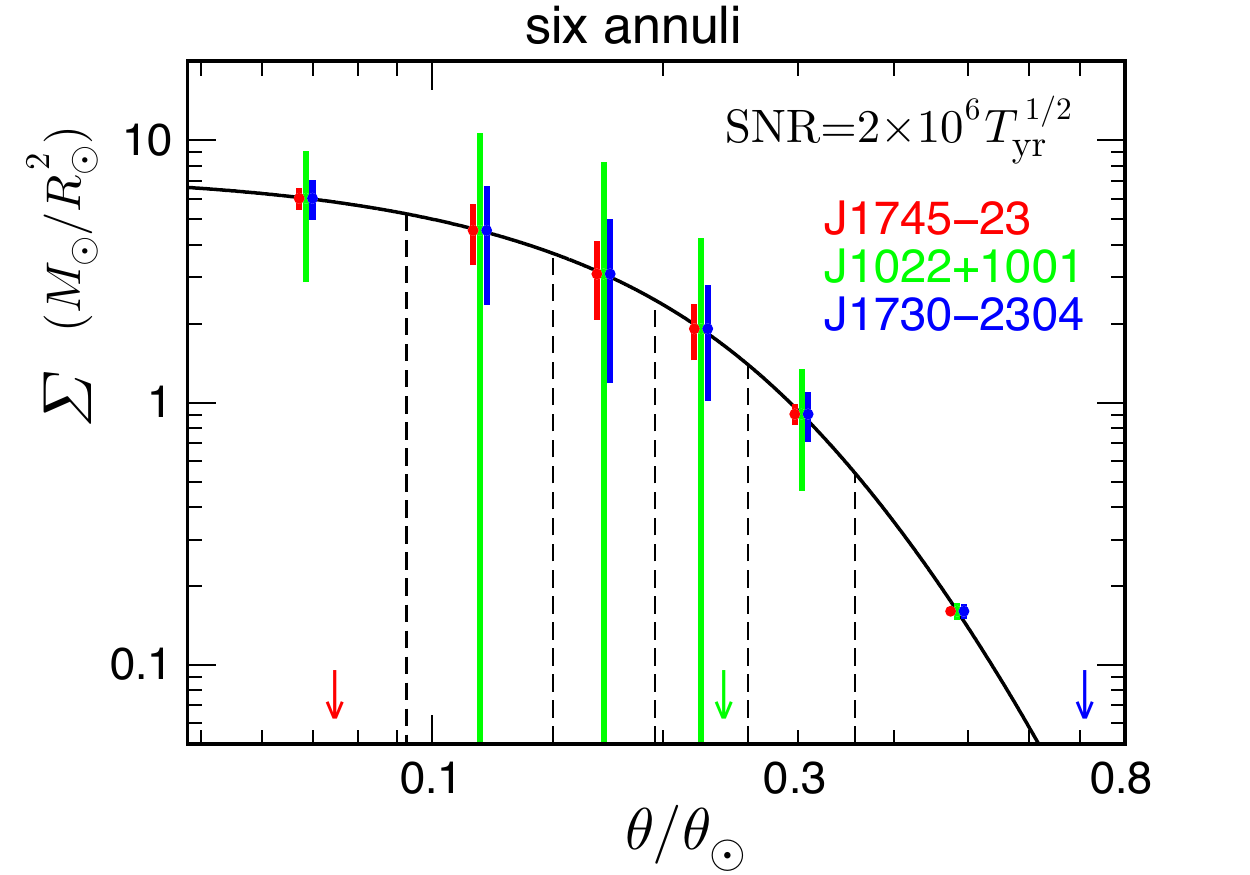}  
\end{minipage}
\begin{minipage}{\columnwidth}
\includegraphics[width=1.2\columnwidth]{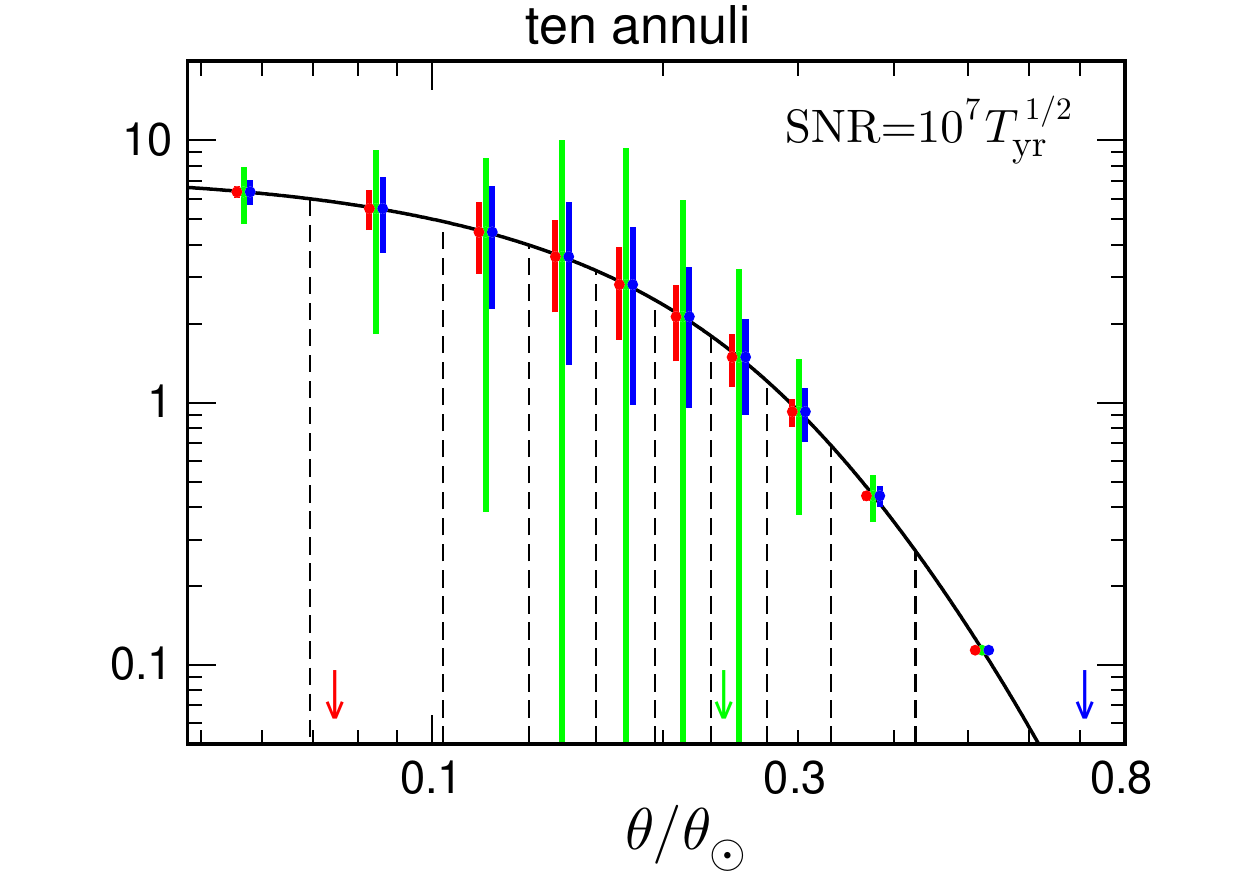}  
\end{minipage}
\caption{Measurement accuracies of the projected solar density as functions of radius. Solid black curve is the fiducial solar model and error bars are the accuracies for each pulsar. In the left and right panels, the density is divided into six and ten annuli in Case (I) with ${\rm SNR}=2 \times 10^6 \, (T/{\rm yr})^{1/2}$ and $10^7 \, (T/{\rm yr})^{1/2}$, respectively. 
Vertical dashed lines are the boundaries between the annuli. 
Down arrows indicate the ecliptic latitude of each pulsar.}
\label{fig_A1-A6}
\end{figure*}


The solar density can (in principle) be plotted as a function of radius in the limit of large $N$ (but a very high SNR is needed).
As a demonstration, Fig.~\ref{fig_A1-A6} plots the measurement accuracies for $N=6$ and $10$.
Here, we follow the same procedure as done for $N=2$ and $3$.
As expected, J1745-23 gives the best accuracy.
Though J1730-2304 has a large impact parameter (denoted by the down blue arrow), it can constrain the inner density due to the diffraction. 
The density of the outermost annulus is determined most precisely, because it has the largest area and the pulsars remain there for the longest time.
The density of the intermediate annuli ($0.1 \lesssim \theta/\theta_\odot \lesssim 0.3$) is poorly determined, because the errors of nearby annuli are strongly correlated.  
For instance, the cross-correlation coefficient of the accuracies in between the i- and j-th annuli, $r_{\rm ij} \equiv (\Gamma^{-1})_{A_{\Sigma {\rm i}} A_{\Sigma {\rm j}}}/[(\Gamma^{-1})_{A_{\Sigma {\rm i}} A_{\Sigma {\rm i}}} (\Gamma^{-1})_{A_{\Sigma {\rm j}} A_{\Sigma {\rm j}}}]^{1/2}$, is large for a close pair of the annuli, $0.6<|r_{\rm ij}|<1$ for $|i-j|=1$.

The strain sensitivities of the planned detectors of the Cosmic Explorer (CE)\footnote{\url{https://cosmicexplorer.org/}} and the Einstein Telescope (ET)\footnote{\url{https://www.et-gw.eu/}} will be approximately $100$ times better than those of the LIGO O3 run. 
The current upper limits in Table~\ref{table_pulsars} were determined from observations taken over approximately one year in the O3 run.
These limits correspond to an upper limit of ${\rm SNR} \lesssim 10$~\citep[Appendix C of ][]{Abbott2019}.
Therefore, in CE and ET observations, the best SNR for these known pulsars is estimated as $\approx 10^3 \, (T/{\rm yr})^{1/2}$~\citep[a similar discussion is given in Section V of][]{Marchant2020}.
Judging from these estimates, measuring the density profile using these pulsars will be a difficult task in the near future.
To alleviate this problem, we require more pulsars than those detected by radio telescopes to date (as discussed in the next section).
Moreover, these pulsars must have high frequencies and small impact parameters. 

\if{}
\section{Compact binary Coalescences}
\label{sec:CBC}

In this section, we consider compact binary coalescence (CBC) signal, whose source location is close to the location of the Sun. 

\subsection{Configurations}

Detector configuration, injection set, sampler settings

\subsection{Results}
\fi{}

\vskip\baselineskip
\vskip\baselineskip

\section{Millisecond pulsars}
\label{sec:MSPs}

\begin{deluxetable}{lrrcc}
\tablecaption{Globular clusters crossing behind the Sun}
\tablehead{
\colhead{name} & \colhead{$\theta_{\rm {\scriptscriptstyle LAT}}/\theta_\odot$} & \colhead{$\theta_{\rm {\scriptscriptstyle h}}/\theta_\odot$} & \colhead{$D_{\rm s}\,$(kpc)} & \colhead{mass ($M_\odot$)}
}
\startdata
NGC 6287 & $0.49$~ & $0.046$ & $7.9$ & $1.5 \times 10^5$ \\
NGC 6717 & $0.50$~ & $0.043$ & $7.5$ & $3.6 \times 10^4$ \\
NGC 6642 & $-0.89$~ & $0.046$ & $8.1$ & $3.4 \times 10^4$
\enddata
\tablecomments{Listed are the ecliptic latitude of the cluster center $\theta_{\rm {\scriptscriptstyle LAT}}$, the half-light angular radius $\theta_{\rm {\scriptscriptstyle h}}$, the distance $D_{\rm s}$, and the mass. Here, the distance and mass are taken from Baumgardt's catalog of globular clusters\footnote{\url{https://people.smp.uq.edu.au/HolgerBaumgardt/globular/}}~\citep[e.g.,][]{Baumgardt2018}.}
\label{table_GC}
\end{deluxetable}

This section roughly estimates the number of MSPs moving behind the Sun that will be found in the near future.

The present ATNF catalog lists approximately $3300$ pulsars, including around $500$ highly spinning pulsars with a spin period of $<10\,{\rm ms}$ (corresponding to $f>200  \, {\rm Hz}$).
These pulsars are the so-called MSPs.
The Milky Way is estimated to host $40000$--$90000$ MSPs~\citep[e.g.,][]{BR2022}.
As the sky fraction of the zodiac belt swept by the Sun is $2\pi \times 2\theta_\odot / (4\pi) \simeq 0.5\%$, the belt probably holds $200$--$400$ MSPs.
Almost none of these MSPs have been found, but tens of them will be identified in ongoing and upcoming radio surveys.
For instance, the Canadian Hydrogen Intensity Mapping Experiment (CHIME)\footnote{\url{https://chime-experiment.ca/en}}, MeerKAT\footnote{\url{https://www.sarao.ac.za/gallery/meerkat/}}, and the Five-hundred-meter Aperture Spherical radio Telescope (FAST)\footnote{\url{https://fast.bao.ac.cn}} will detect $\approx 1000$ MSPs over the sky~\citep[][their Table 1]{Lorimer2019}.
These performances will be surpassed by the Square Kilometer Array (SKA) Phase 1, which will detect $1500$ MSPs~\citep{Keane2015}.

At present, approximately $220$ MSPs have been found in $33$ globular clusters\footnote{\url{https://www3.mpifr-bonn.mpg.de/staff/pfreire/GCpsr.html}}; 
$\sim$ seven MSPs per cluster on average.
Per unit mass, pulsars are two or three orders of magnitude more abundant in clusters than in the Galactic disk~\citep{Freire2013}.
Therefore, many MSPs probably exist in clusters.
Among the $157$ clusters listed in the McMaster catalog of Milky Way globular clusters\footnote{Data may be downloaded from \url{https://physics.mcmaster.ca/Fac\_Harris/mwgc.dat}}~\citep[version 2010,][]{Harris1996,Harris2010}, three globular clusters move behind the Sun.
The data are summarized in Table ~\ref{table_GC}.
Although no MSPs have been discovered in these clusters\footnote{\url{https://www3.mpifr-bonn.mpg.de/staff/pfreire/GCpsr.html}},
tens of MSPs are expected in each cluster.

\section{Lensing by a Galactic star}
\label{sec:discussion}

This section briefly discusses GW lensing by a Milky Way star along the line of sight to the pulsar. 
As a star crosses in front of the pulsar, lensing imprints a time-dependent modulation on the waveform.
This event might be confused with the solar lensing.
We comment that these two events can be distinguished due to the following reasons.
First, the stellar lensing is observed only once, but the solar lensing is observed every year on the same day. 
In addition, the probability of the stellar lensing is not high, $\approx 10^{-6}$ ${\rm yr}^{-1}$, on average over the full sky ~\citep[e.g.,][]{Han2008}.
If the pulsar is in a globular cluster, the probability of lensing by a cluster star (so-called self-lensing) can reach $\approx 0.1$ ${\rm yr}^{-1}$~\citep{Kiroglu2022}.
Second, the lensing modulation by the star (modeled as a point mass) differs from the solar modulation~\citep[e.g., Figs. 3 and 5 of][]{Marchant2020}. For example, the typical event duration is $\approx 10$ days~\citep[e.g.,][]{SKW2006}, which is much longer than the solar lensing duration $\approx$ half a day.
Third, if radio observations are available for the stellar lensing, double signals (with different amplitudes and arrival times) from the single pulsar may be detected.

\section{Conclusions}
\label{sec:conclusion}

We have studied the detectability of solar density profiling using GW lensing with known pulsars.
After selecting suitable pulsars with high frequencies and small impact parameters to the Sun from the ATNF catalog, we calculated the measurement accuracy of the overall amplitude of the solar gravitational potential using a Fisher analysis.
The lensing signature can be detected with $3 \sigma$ confidence when ${\rm SNR} \approx 100 \, (f/300 \, {\rm Hz})^{-1}$ during one year of observation (Table~\ref{table_Apsi} and Eq.~(\ref{deltaApsi})).
The detection is therefore improved with high-frequency pulsars. 
The signature can be detected even if the pulsar trajectory does not pass behind the Sun (in such cases, the impact parameter to the Sun does not significantly affect the detectability). 
We found that the detectability is degraded by parameter degeneracy with the constant phase ($\phi_0$ in Eq.~(\ref{lensed_waveform})) of the waveform.
Next, we divided the projected density profile into $N$ annuli ($N=2,3,\cdots$) and calculated the measurement accuracy of each annulus mass. 
If three known pulsars move behind the Sun with $f>100 \, {\rm Hz}$, a high SNR ($\gtrsim 10^4$) in a year-long observation is required for measuring the density profile in the two-annulus case (Fig.~\ref{fig_A1A2_error}).
The SNR must be raised to $\gtrsim 10^5$ in the three-annulus case (Fig.~\ref{fig_A1A2A3_error}) and $\gtrsim 10^6$ in the six-annulus case (Fig.~\ref{fig_A1-A6}).

Currently, only two known MSPs move behind the Sun, but $200$--$400$ can potentially exist within our Galaxy (Section \ref{sec:MSPs}).
Ongoing and future radio surveys such as SKA, CHIME, and FAST will find tens of these MSPs in the coming decades.

The solar density profile has already been measured to percent-level precision based on global helioseismology~\citep[see the recent review by][]{Basu2016}. 
GW lensing will likely become a complementary and independent method for probing the Sun in the future.

\section*{Acknowledgments}
We thank Hideyuki Hotta for his useful comment on helioseismology. 
We thank the CASS (CSIRO (Commonwealth Scientific and Industrial Research Organisation) Astronomy and Space Science) pulsar group for developing and maintaining the ATNF pulsar catalog.
This work is supported in part by MEXT Japan and JSPS KAKENHI Grant Numbers of JP22H00130 (RT), JP20H05855 (RT), JP17H06359 (TS), JP21H05453 (TS), and JP19K03864 (TS). 


\appendix

\section{Validity of the thin-lens approximation}
\label{sec:systematics}

\cite{Suyama2005} investigated the validity of the thin-lens approximation in wave optics.
They considered plane waves entering a lens and 
numerically solved the wave equation to obtain the exact lensed waveform (without the thin-lens approximation).
They considered three spherical lens models (a uniform density sphere, a singular isothermal sphere truncated at a finite radius, and the Hernquist profile). 
They found that the relative error of $F$ is smaller than the lens radius over $D_{\rm L}$ and the maximum error is achieved when the wavelength is comparable to the Schwarzschild radius of the lens.
This suggests that the error of $F$ is $\lesssim R_\odot/(1 \, {\rm{au}}) \simeq 0.5 \, \%$ in our case.


\bibliographystyle{aasjournal}
\bibliography{refs}


\end{document}